\documentclass[11pt,a4paper]{article}
\pdfoutput=1
\usepackage{jheppub}
\usepackage{amsmath}
\usepackage{amssymb}
\usepackage{graphics}
\usepackage[active]{srcltx}
\usepackage{pdfsync}
\usepackage{shuffle}
\usepackage{slashed}

\usepackage{subfigure}

\newcommand{\bea}{\begin{eqnarray}}
\newcommand{\eea}{\end{eqnarray}}
\newcommand{\nn}{\nonumber \\}

\def\W #1{\widetilde{#1}}

\def\Tr{\mathop{\rm Tr}}

\def\eref#1{(\ref{#1})}

\def\a{{\alpha}}

\def\b{{\beta}}

\title{Note on tree NLSM amplitudes and soft theorems}
\author[a]{Kang Zhou,}
\author[b]{Fang-Stars Wei,}

\affiliation[a]{Center for Gravitation and Cosmology, College of Physical Science and Technology, Yangzhou University,\\
No.180, Siwangting Road, Yangzhou, 225009, P.R. China.}
\affiliation[b]{Center for Gravitation and Cosmology, College of Physical Science and Technology, Yangzhou University,\\
No.180, Siwangting Road, Yangzhou, 225009, P.R. China.}

\emailAdd{zhoukang@yzu.edu.cn}
\emailAdd{mx120220339@stu.yzu.edu.cn}

\date{\today}
\abstract{We use the universality of single soft behavior, together with the double copy structure, to completely determine the tree amplitudes of non-linear sigma model (NLSM). We first figure out the Adler's zero for $4$-point NLSM amplitudes, by considering kinematics. Then, we assume the universality of the Adler's zero, and use this requirement to construct general tree NLSM amplitudes in the expanded formula, i.e., the formula of expanding NLSM amplitudes to bi-adjoint scalar amplitudes. We also derive double soft factors for tree NLSM amplitudes, based on the resulting expanded formula.
}

\begin{document}

\maketitle \flushbottom

\section{Introduction}
\label{sec-intro}

Soft theorems describe the universal behavior of scattering amplitudes when one or more external massless momenta are taken to near zero. This limit can be achieved by re-scaling
the massless momenta via a soft parameter as $k^\mu\to \tau k^\mu$, and taking the limit $\tau\to 0$. Soft theorems then state the factorization
of amplitudes. For instance, when one of external gravitons is taken to be soft, the $(n+1)$-point GR amplitude factorizes as \cite{Cachazo:2014fwa,Schwab:2014xua,Afkhami-Jeddi:2014fia}
\bea
{\cal A}_{n+1}\,\to\,\Big(\tau^{-1}\,S^{(0)}_h+\tau^0\,S^{(1)}_h+\tau\,S^{(2)}_h+\cdots\Big)\,{\cal A}_n\,,~~~~\label{softtheo}
\eea
where ${\cal A}_n$ is the sub-amplitude of ${\cal A}_{n+1}$, which is generated from ${\cal A}_{n+1}$ by removing the soft external graviton.
The universal operators $S^{(0)}_h$, $S^{(1)}_h$, $S^{(2)}_h$ are called soft factors, or soft operators, at leading, sub-leading, and sub-sub-leading orders.

Soft theorems were exploited in the construction of tree amplitudes, such as the related on-shell recursion relations, and the inverse soft theorem program, and so on \cite{Cheung:2014dqa,Luo:2015tat,Cheung:2018oki,Elvang:2018dco,Cachazo:2016njl,Rodina:2018pcb,Boucher-Veronneau:2011rwd,Nguyen:2009jk}. Impressively, in \cite{Rodina:2018pcb}, it was shown that soft theorems uniquely fix tree amplitudes. However, the previous constructions mentioned above require the explicit forms of soft factors. To derive soft factors, one need a certain expression for amplitudes, such as summations of contributions from Feynman diagrams \cite{Low,Weinberg}, Britto-Cachzo-Feng-Witten (BCFW) on-shell recursion relations \cite{Cachazo:2014fwa,Casali:2014xpa,Britto:2004ap,Britto:2005fq}, Cachazo-He-Yuan (CHY) contour integrals \cite{Schwab:2014xua,Afkhami-Jeddi:2014fia,Cachazo:2013gna,Cachazo:2013hca, Cachazo:2013iea, Cachazo:2014nsa,Cachazo:2014xea}, and so on.
Then a logical confusion is caused, the tree amplitude, and the soft factor, which one determines another one?

On the other hand, the factorization in \eref{softtheo} has a natural physical picture. Roughly speaking, in the soft limit the soft particle can be thought as vanishing, leaving a lower-point amplitude with the soft external leg removed, and the universal soft factors carried by the soft particle. Thus, to avoid the logical confusion mentioned previously, it is natural to ask, can we take the soft behavior in \eref{softtheo} as the principle, and use it to construct tree amplitudes, without knowing the explicit forms of soft factors? In the recent work of one of two authors for the present note, it was shown that such construction can be realized at least for tree amplitudes of Yang-Mills-scalar, pure Yang-Mills, Einstein-Yang-Mills and pure gravitational theories \cite{Zhou:2022orv}. The factorization behaviors, together with assuming the universality of soft factors, and the double copy structure \cite{Kawai:1985xq,Bern:2008qj,Chiodaroli:2014xia,Johansson:2015oia,Johansson:2019dnu}, completely determine tree amplitudes of above theories.

This note is the continuation of the previous work in \cite{Zhou:2022orv}, this time applying the similar idea to consider the tree amplitudes of nonlinear sigma model (NLSM). The tree NLSM amplitudes vanishing in the single soft limit (one of external leg being soft), known as Adler's zero
\cite{Adler}. For such case, the factorization behavior in \eref{softtheo} does not occur. However, one can still talk about the universality of the single soft behavior. As will be seen in subsection.\ref{subsec-34p} in section.\ref{sec-expan}, it is easy to figure out the Adler's zero for the $4$-point tree NLSM amplitude, by considering kinematics. Then, the universality of soft behavior indicates that all tree NLSM amplitudes vanish in the single soft limit.
The assumption of universality, together with other physical conditions such as double copy structure and manifest permutation invariance among external legs, uniquely fix the general tree NLSM amplitudes, in the expanded formula, i.e., the formula of expanding NLSM amplitudes to bi-adjoint scalar (BAS) amplitudes. We emphasize that the Adler's zero was not assumed at the beginning, it was deduced. Using the resulting expanded formula, we also derive double soft factors for the tree NLSM amplitudes \cite{Cachazo:2015ksa,Du:2015esa}, which describe the behavior when two external scalars are taken to be soft simultaneously.

The remainder of this note is organized as follows. In section.\ref{sec-background}, we give a quick introduction to the necessary background, including BAS tree amplitudes, and expansions of other amplitudes to them. In section.\ref{sec-expan}, we use the the universality of single soft behavior, to construct the tree NLSM amplitudes in the expanded formula. In section.\ref{sec-double-soft}, we derive the double soft factor for tree NLSM amplitudes, based on the expanded formula obtained in section.\ref{sec-expan}. Finally, we close with a brief summery in section.\ref{sec-conclu}.

\section{Background}
\label{sec-background}

In this section we rapidly review the necessary background. In subsection.\ref{subsecBAS}, we introduce the tree level amplitudes of bi-adjoint scalar (BAS) theory. Some notations and technics which will be used in subsequent sections are also included. In subsection.\ref{subsecexpand}, we rapidly review the expansions of tree amplitudes to BAS amplitudes, including the choice of basis, as well as the double copy structure for coefficients.

\subsection{Tree level BAS amplitudes}
\label{subsecBAS}

The BAS theory describes the bi-adjoint scalar field $\phi_{a\bar{a}}$ with the Lagrangian
\bea
{\cal L}_{\rm BAS}={1\over2}\,\partial_\mu\phi^{a\bar{a}}\,\partial^{\mu}\phi^{a\bar{a}}-{\lambda\over3!}\,f^{abc}f^{\bar{a}\bar{b}\bar{c}}\,
\phi^{a\bar{a}}\phi^{b\bar{b}}\phi^{c\bar{c}}\,,~~\label{Lag-BAS}
\eea
where the structure constant $f^{abc}$ and generator $T^a$ satisfy
\bea
[T^a,T^b]=if^{abc}T^c\,,
\eea
and the dual color algebra encoded by $f^{\bar{a}\bar{b}\bar{c}}$ and $T^{\bar{a}}$ is analogous.
The tree level amplitudes of this theory contain only propagators, and can be decomposed into double color ordered partial amplitudes via the standard technic.
Each double color ordered partial amplitude is simultaneously planar with respect to two color orderings, arise from expanding the full $n$-point amplitude to $\Tr(T^{a_{\sigma_1}}\cdots T^{a_{\sigma_n}})$ and $\Tr(T^{\bar{a}_{\bar{\sigma}_1}}\cdots T^{\bar{a}_{\bar{\sigma}_n}})$ respectively,
where $\sigma_i$ and $\bar{\sigma}_i$ denote permutations among all external scalars.

\begin{figure}
  \centering
  \includegraphics[width=6cm]{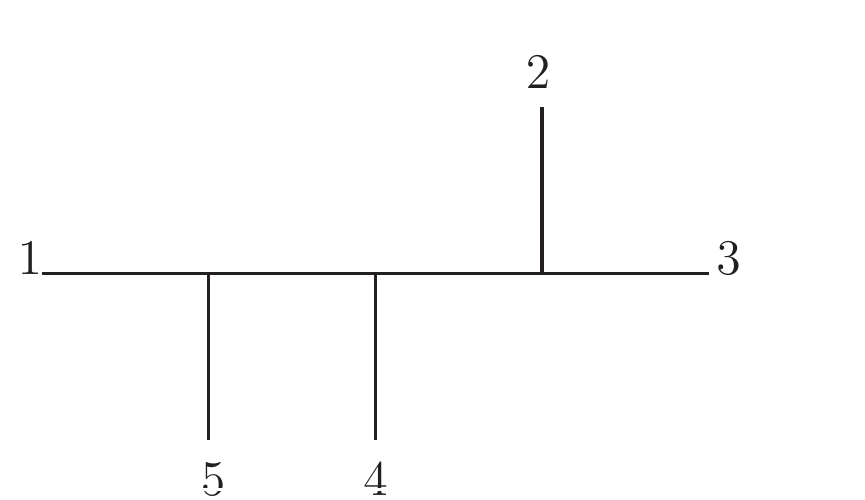}
   \includegraphics[width=6cm]{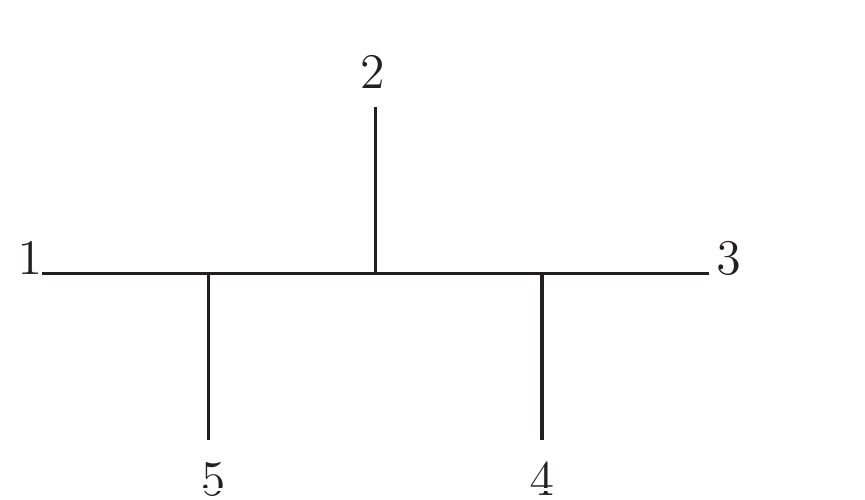}  \\
  \caption{Two $5$-point diagrams}\label{5p}
\end{figure}

\begin{figure}
  \centering
   \includegraphics[width=4cm]{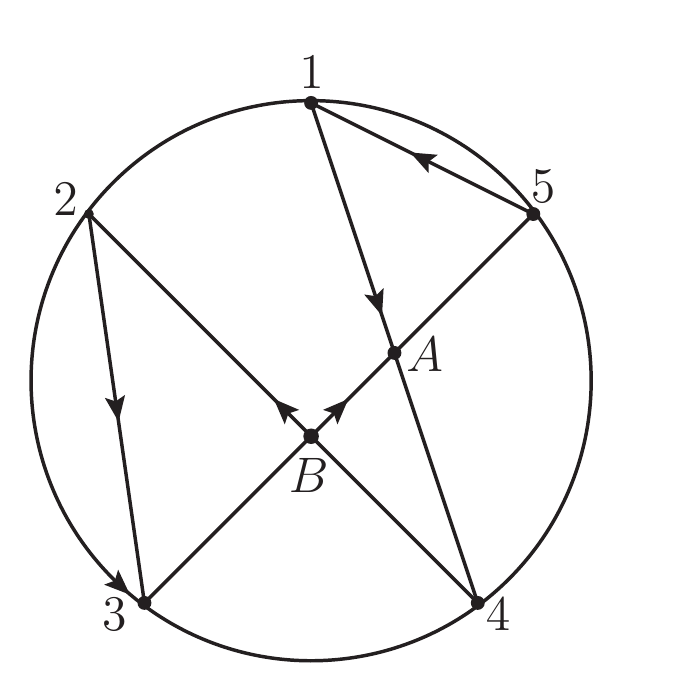}
   \includegraphics[width=4cm]{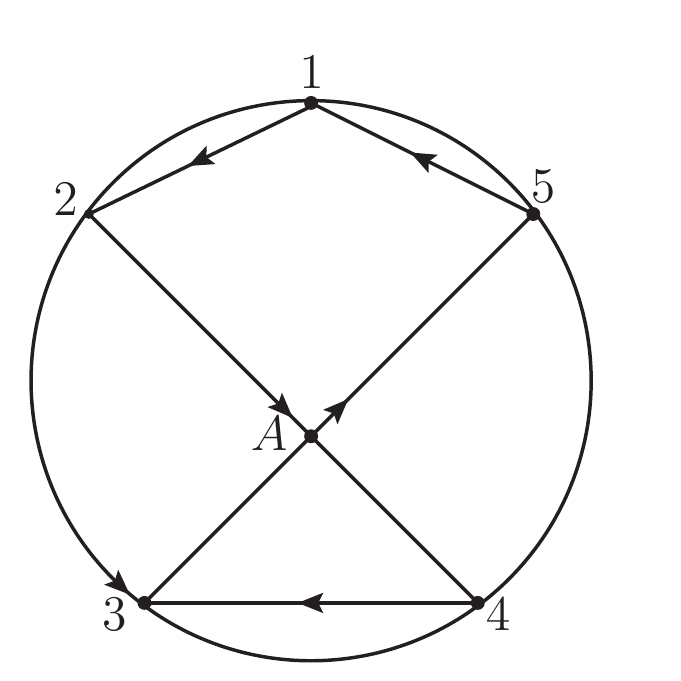} \\
  \caption{Diagram for ${\cal A}_S(1,2,3,4,5|1,4,2,3,5)$ and ${\cal A}_S(1,2,3,4,5|1,2,4,3,5)$.}\label{dis14235}
\end{figure}

To calculate double color ordered partial amplitudes, it is convenient to employ the diagrammatical method proposed by Cachazo, He and Yuan in \cite{Cachazo:2013iea}.
For a BAS amplitude whose double color-orderings are given, this method provides the corresponding Feynman diagrams as well as the overall sign directly.
Let us consider the $5$-point example ${\cal A}_{\rm S}(1,2,3,4,5|1,4,2,3,5)$.
In Figure.\ref{5p}, the first diagram satisfies both two color orderings $(1,2,3,4,5)$ and $(1,4,2,3,5)$, while the second one satisfies the ordering
$(1,2,3,4,5)$ but not $(1,4,2,3,5)$. Thus, the first diagram is allowed by the double color orderings $(1,2,3,4,5|1,4,2,3,5)$, while the second one is not. It is easy to see other diagrams are also forbidden by the ordering $(1,4,2,3,5)$, thus the first diagram in Figure.\ref{5p} is the only diagram contributes to the amplitude
${\cal A}_{\rm S}(1,2,3,4,5|1,4,2,3,5)$.
The Feynman diagrams contribute to a given BAS amplitude can be obtained via systematic diagrammatical rules. For the above example, one can draw a disk diagram as follows.
\begin{itemize}
\item Draw points on the boundary of the disk according to the first ordering $(1,2,3,4,5)$.
\item Draw a loop of line segments which connecting the points according to the second ordering $(1,4,2,3,5)$.
\end{itemize}
The obtained disk diagram is shown in the first diagram in Figure.\ref{dis14235}. From the diagram, one can see that two orderings share the boundaries $\{1,5\}$ and $\{2,3\}$. These co-boundaries
indicate channels ${1/s_{15}}$ and ${1/s_{23}}$, therefore the first Feynman diagram in Figure.\ref{5p}. Then the BAS amplitude ${\cal A}_{\rm S}(1,2,3,4,5|1,4,2,3,5)$ can be computed as
\bea
{\cal A}_{\rm S}(1,2,3,4,5|1,4,2,3,5)={1\over s_{23}}{1\over s_{51}}\,,
\eea
up to an overall sign. The Mandelstam variable $s_{i\cdots j}$ is defined as
\bea
s_{i\cdots j}\equiv K_{i\cdots j}^2\,,~~~~K_{i\cdots j}\equiv\sum_{a=i}^j\,k_a\,,~~~~\label{mandelstam}
\eea
where $k_a$ is the momentum carried by the external leg $a$.

As another example, let us consider the BAS amplitude ${\cal A}_{\rm S}(1,2,3,4,5|1,2,4,3,5)$. The corresponding disk diagram is shown in the second configuration in
Figure.\ref{dis14235}, and one can see two orderings have co-boundaries $\{3,4\}$ and $\{5,1,2\}$. The co-boundary $\{3,4\}$ indicates the channel ${1/ s_{34}}$. The co-boundary $\{5,1,2\}$ indicates the channel ${1/s_{512}}$ which is equivalent to $1/s_{34}$, as well as sub-channels ${1/ s_{12}}$ and ${1/ s_{51}}$. Using the above decomposition, one can calculate ${\cal A}_{\rm S}(1,2,3,4,5|1,2,4,3,5)$ as
\bea
{\cal A}_{\rm S}(1,2,3,4,5|1,2,4,3,5)={1\over s_{34}}\Big({1\over s_{12}}+{1\over s_{51}}\Big)\,,
\eea
up to an overall sign.

The overall sign, determined by the color algebra, can be fixed by following rules.
\begin{itemize}
\item Each polygon with odd number of vertices contributes
a plus sign if its orientation is the same as that of the disk and a minus sign if opposite.
\item Each polygon with even number of vertices always contributes a minus sign.
\item Each intersection point contributes a minus sign.
\end{itemize}
We now apply these rules to previous examples. In the first diagram in Figure.\ref{dis14235}, the polygons are three triangles, namely $51A$, $A4B$ and $B23$, which contribute $+$, $-$, $+$ respectively, while two intersection points $A$ and $B$ contribute two $-$. In the second one in Figure.\ref{dis14235}, the polygons are $512A$ and $A43$, which contribute two $-$, while the intersection point $A$ contributes $-$. Then we arrive at the full results
\bea
{\cal A}_{\rm S}(1,2,3,4,5|1,4,2,3,5)&=&-{1\over s_{23}}{1\over s_{51}}\,,\nn
{\cal A}_{\rm S}(1,2,3,4,5|1,2,4,3,5)&=&-{1\over s_{34}}\Big({1\over s_{12}}+{1\over s_{51}}\Big)\,.
\eea

\begin{figure}
  \centering
   \includegraphics[width=4cm]{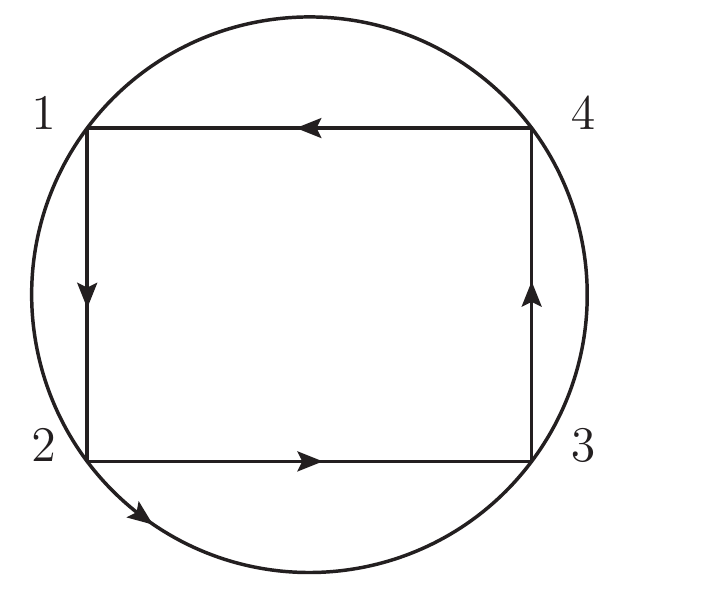} \\
  \caption{The overall sign $+$ under the new convention.}\label{newconvention}
\end{figure}

In the reminder of this note, we adopt another convention for the overall sign. If the line segments form a convex polygon, and the orientation of the convex polygon is the same as that of the disk, then the overall sign is $+$. For instance, the disk diagram in Figure.\ref{newconvention} indicates the overall sign $+$ under the new convention, while the old convention gives $-$ according to the square formed by four line segments. Notice that the diagrammatical rules described previously still give the related sign between different disk diagrams. For example, two disk diagrams in Figure.\ref{dis14235} shows that the relative sign between ${\cal A}_{\rm S}(1,2,3,4,5|1,4,2,3,5)$ and
${\cal A}_{\rm S}(1,2,3,4,5|1,2,4,3,5)$ is $+$. The advantage of the new convention is that when removing a soft external scalar, the resulting sub-amplitude carries the same sign as the original one.

When considering the soft limit, the $2$-point channels play the central role. Since the partial BAS amplitude carries two color orderings, if the $2$-point channel contributes $1/s_{ab}$ to the amplitude, external legs $a$ and $b$ must be adjacent to each other in both two orderings.
Suppose the first color ordering is $(\cdots,a,b,\cdots)$, then $1/s_{ab}$ is allowed by this ordering. To denote if it is allowed by another one, we introduce the symbol $\delta_{ab}$ whose ordering of two subscripts $a$ and $b$ is determined by the first color ordering\footnote{The Kronecker symbol will not appear in this paper, thus we hope the notation $\delta_{ab}$ will not confuse the readers.}. The value of $\delta_{ab}$ is $\delta_{ab}=1$ if another color ordering is $(\cdots,a,b,\cdots)$, $\delta_{ab}=-1$ if another color ordering is $(\cdots,b,a,\cdots)$, due to the ani-symmetry of the structure constant, i.e., $f^{abc}=-f^{bac}$, and $\delta_{ab}=0$
otherwise. From the definition, it is straightforward to see $\delta_{ab}=-\delta_{ba}$, and a simple but useful identity
\bea
\sum_{b\neq a}\,\delta_{ab}=0\,.~~\label{iden}
\eea

Before ending this subsection, we discuss the single soft behavior of BAS amplitudes at the leading order. Consider the double color ordered BAS amplitude ${\cal A}_{\rm S}(1,\cdots,n|\sigma_n)$. We re-scale $k_i$
as $k_i\to\tau k_i$, and expand the amplitude in $\tau$. The leading order contribution manifestly aries from $2$-point channels $1/s_{1(i+1)}$
and $1/ s_{(i-1)i}$ which provide the $1/\tau$ order contributions, namely,
\bea
{\cal A}^{(0)}_{\rm S}(1,\cdots,n|\sigma_n)&=&{1\over \tau}\Big({\delta_{i(i+1)}\over s_{i(i+1)}}+{\delta_{(i-1)i}\over s_{(i-1)i}}\Big)\,
{\cal A}_{\rm S}(1,\cdots,i-1,\not{i},i+1,\cdots,n|\sigma_n\setminus i)\nn
&=&S^{(0)}_s(i)\,{\cal A}_{\rm S}(1,\cdots,i-1,\not{i},i+1,\cdots,n|\sigma_n\setminus i)\,,~~~\label{soft-theo-s}
\eea
where $\not{i}$ stands for removing the leg $i$, $\sigma_n\setminus1$ means the color ordering generated from $\sigma_n$ by eliminating $i$. The leading soft operator $S^{(0)}_s(i)$ for the scalar $i$ is extracted as
\bea
S^{(0)}_s(i)={1\over \tau}\,\Big({\delta_{i(i+1)}\over s_{i(i+1)}}+{\delta_{(i-1)i}\over s_{(i-1)i}}\Big)\,,~~~~\label{soft-fac-s-0}
\eea
which acts on external scalars which are adjacent to $i$ in two color orderings.

\subsection{Expanding tree level amplitudes to BAS basis}
\label{subsecexpand}

\begin{figure}
  \centering
   \includegraphics[width=4cm]{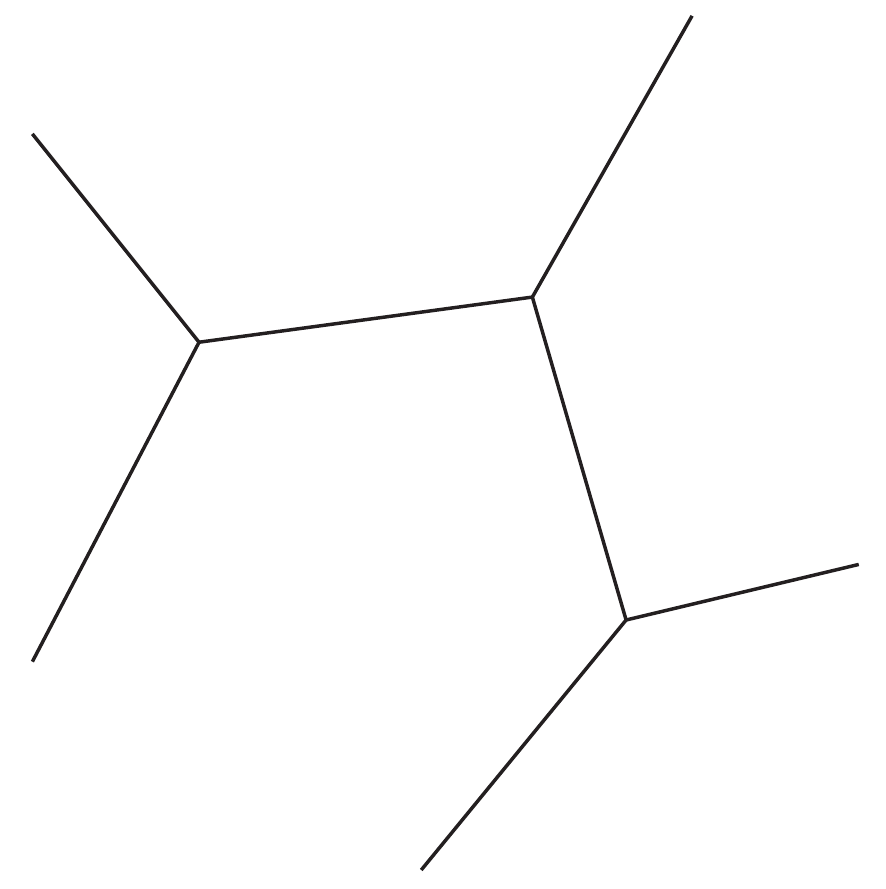}
   \includegraphics[width=4cm]{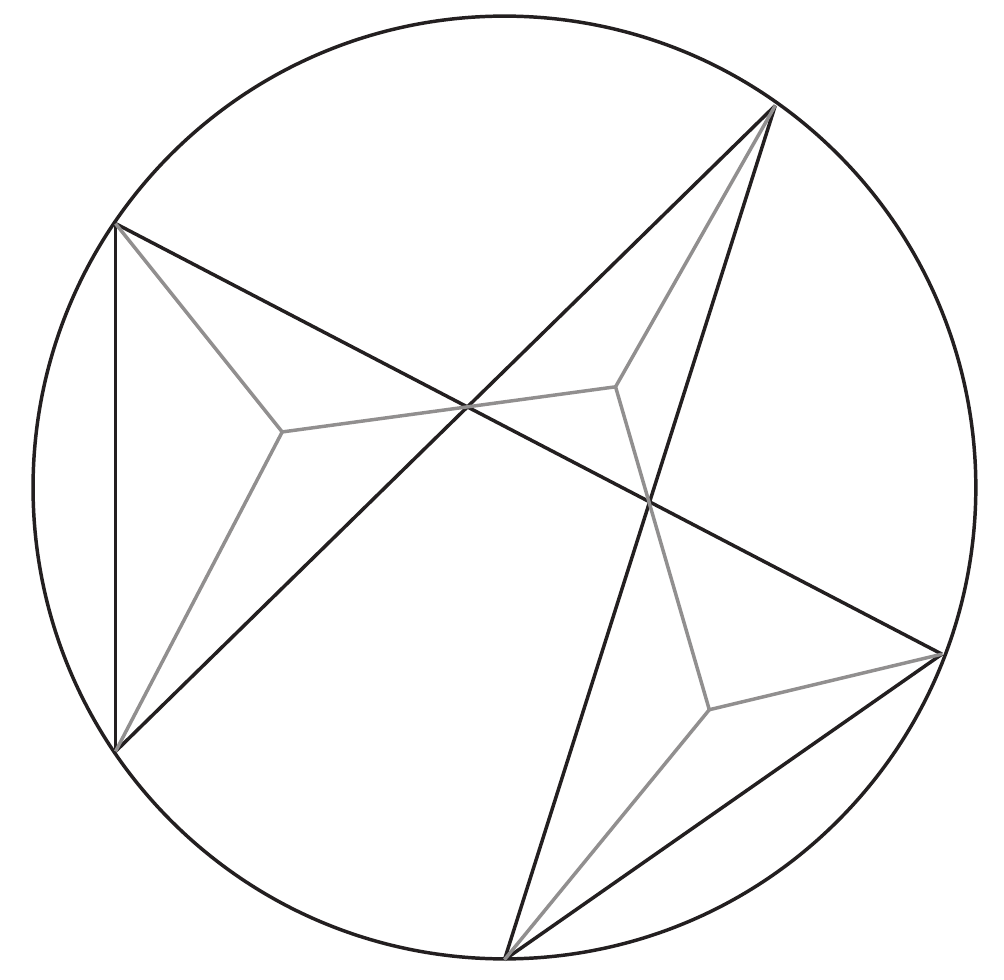} \\
  \caption{Map between Feynman diagram and disk diagram.}\label{example}
\end{figure}

Tree level amplitudes which contain only massless particles can be expanded to double color ordered BAS amplitudes,
due to the observation that each Feynman diagram for pure propagators can be mapped to at least one disk diagram whose polygons are all triangles. An illustrative example is given in Figure.\ref{example}. Since each tree amplitude can be expanded by tree Feynman diagrams, and each Feynman diagram contributes propagators together with a numerator without any pole, one can conclude that each tree amplitude can be expanded to double color ordered BAS amplitudes, with coefficients which are polynomials depend on Lorentz invariants created by external kinematical variables.
To realize the expansion, one need to find the basis consists of BAS amplitudes.
Such basis can be determined by the well known Kleiss-Kuijf (KK) relation \cite{Kleiss:1988ne}
\bea
{\cal A}_{\rm S}(1,\vec{\pmb{\a}},n,\vec{\pmb{\b}}|\sigma_n)=(-)^{|\pmb{\a}|}\,{\cal A}_{\rm S}(1,\vec{\pmb{\a}}\shuffle\vec{\pmb{\b}}^T,n|\sigma_n)\,.~~~\label{KK}
\eea
Here $\vec{\pmb{\a}}$ and $\vec{\pmb{\b}}$ are two ordered subsets of external scalars, and $\vec{\pmb{\b}}^T$ stands for the ordered set generated from $\vec{\pmb{\b}}$ by reversing the original ordering. The $n$-point BAS amplitude ${\cal A}_{\rm S}(1,\vec{\pmb{\a}},n,\vec{\pmb{\b}}|\sigma_n)$ at the l.h.s of \eref{KK} carries two color orderings, one is $(1,\vec{\pmb{\a}},n,\vec{\pmb{\b}})$, another one is denoted by $\sigma_n$. The symbol $\shuffle$ means summing over all
possible shuffles of two ordered sets $\vec{\pmb{\b}}_1$ and $\vec{\pmb{\b}}_2$, i.e., all permutations in the set $\vec{\pmb{\b}}_1\cup \vec{\pmb{\b}}_2$ while preserving the orderings
of $\vec{\pmb{\b}}_1$ and $\vec{\pmb{\b}}_2$. For instance, suppose $\vec{\pmb{\b}}_1=\{1,2\}$ and $\vec{\pmb{\b}}_2=\{3,4\}$, then
\bea
\vec{\pmb{\b}}_1\shuffle \vec{\pmb{\b}}_2=\{1,2,3,4\}+\{1,3,2,4\}+\{1,3,4,2\}+\{3,1,2,4\}+\{3,1,4,2\}+\{3,4,1,2\}\,.~~~~\label{shuffle}
\eea
The analogous KK relation holds for another color ordering $\sigma_n$.
The KK relation implies that different double color ordered BAS amplitudes are not independent, thus the basis can be chosen as BAS amplitudes
${\cal A}_{\rm S}(1,\sigma_1,n|1,\sigma_2,n)$, with $1$ and $n$ are fixed at two ends in each color ordering. We call such basis the KK BAS basis. Based on the discussion above, the KK BAS basis can provide any structure of massless propagators, thus any amplitude which includes only massless particles can be expanded to this basis\footnote{The well known Bern-Carrasco-Johansson (BCJ) relation \cite{Bern:2008qj,Chiodaroli:2014xia,Johansson:2015oia,Johansson:2019dnu} links BAS amplitudes in the KK basis together, and the independent BAS amplitudes can be obtained by fixing three legs at three particular positions in the color orderings. However, in the BCJ relation, coefficients of BAS amplitudes depend on Mandelstam variables, this character leads to poles in coefficients when expanding to BCJ basis. On the other hand, when expanding to KK basis, one can find the expanded formula in which the coefficients contain no pole. In this paper, we choose the KK basis since we hope all poles of tree amplitudes are included in basis, and coefficients only serve as numerators.}.
In other words, the basis provides propagators, and the coefficients in expansions provide numerators. From this point of view, one can regard the BAS KK basis as the complete set of different structures of propagators, and forget the corresponding Lagrangian in \eref{Lag-BAS}.

In this note, we will consider the expansion of NLSM amplitudes. The color ordered NLSM amplitude ${\cal A}_{\rm N}(1,\sigma_{n-2},n)$ can be expanded to KK BAS basis as
\bea
{\cal A}_{\rm N}(1,\sigma_{n-2},n)=\sum_{\sigma'_{n-2}}\,{\cal C}(\sigma'_{n-2},k_i)\,{\cal A}_{\rm S}(1,\sigma'_{n-2},n|1,\sigma_{n-2},n)\,,~~~\label{exp-N-KK}
\eea
where $\sigma_{n-2}$ and $\sigma'_{n-2}$ are permutations among $(n-2)$ external legs in $\{2,\cdots,n-1\}$. The double copy structure \cite{Kawai:1985xq,Bern:2008qj,Chiodaroli:2014xia,Johansson:2015oia,Johansson:2019dnu} indicates that the coefficient ${\cal C}(\sigma'_{n-2},k_i)$ depends on
momenta $k_i$ carried by external scalars, permutations $\sigma'_{n-2}$, but is independent of the permutation $\sigma_{n-2}$\footnote{Originally, the double copy means the GR amplitude can be factorized as ${\cal A}_{G}={\cal A}_Y\times {\cal S}\times{\cal A}_Y$, where the kernel ${\cal S}$ is obtained by inverting propagators. Our assumption that the coefficients depend on only one color ordering is equivalent to the original version, see in \cite{Zhou:2022orv}.}. Thus, suppose we replace $(1,\sigma_{n-2},n)$ by the more general ordering $\sigma_n$ among all external legs, without fixing $1$ and $n$ at any position, the expansion in \eref{exp-N-KK} still holds.
The coefficients ${\cal C}(\sigma'_{n-2},k_i)$ will be constructed in the next section.

\section{Expanded NLSM amplitudes}
\label{sec-expan}

Tree amplitudes can be expressed in various manners, the expansions of tree amplitudes to KK BAS basis discussed in subsection.\ref{subsecexpand} provide one of them. To determine the tree NLSM amplitudes, it is sufficient to fix coefficients ${\cal C}(\sigma'_{n-2},k_i)$
in \eref{exp-N-KK}. This is the goal of the present section.

The standard NLSM Lagrangian in the Cayley parametrization is given as
\bea
{\cal L}_{\rm N}={1\over 8\lambda^2}\,{\rm Tr}(\partial_\mu {\rm U}^\dag\partial^\mu {\rm U})\,,~~\label{Lag-N}
\eea
with
\bea
{\rm U}=(\mathbb{I}+\lambda\Phi)\,(\mathbb{I}-\lambda\Phi)^{-1}\,,
\eea
where $\mathbb{I}$ is the identity matrix, and $\Phi=\phi_IT^I$, with $T^I$ the generators of $U(N)$. Fields $\phi_I$ describe massless scalars, and the accompanied generators $T^I$ indicates the color ordering for the corresponding partial tree amplitudes. From the Lagrangian in \eref{Lag-N}, we see the mass dimension of coupling constant $\lambda$ is $(2-d)/2$, in $d$ dimensional space-time. The mass dimension of the $n$-point amplitude is $d-{d-2\over2}n$, the coupling constants contribute $(2-d)(n-2)/2$, thus the kinematic part must have mass dimension $2$.

Thus, our purpose is to seek the $n$-point amplitude (kinematic part) ${\cal A}_{\rm N}(\sigma_n)$ for pure massless scalars which carry the color ordering $\sigma_n$ among $n$ external scalars, and has mass dimension $2$. As will be seen, the requirements mentioned above, together with the universality of soft behavior, and the permutation symmetry among external scalars, are sufficient to determine the tree NLSM amplitudes completely, without using any other information. In this sense, in the reminder of this section one can forget the traditional Lagrangian and Feynman rules, only concentrate on the color ordering and mass dimension.
Since the propagators contribute the mass dimension $-2(n-3)$, the mass dimension of numerators can be fixed as $2(n-2)$.
This is the mass dimension of coefficients ${\cal C}(\sigma'_{n-2},k_i)$ in the expansion \eref{exp-N-KK}, which plays the important role in the subsequent subsections.

We will argue that the $3$-point NLSM tree amplitude does not exist, while the $4$-point one has the vanishing single soft behavior at the $\tau^0$ order, via the general consideration of  Mandelstam variables. Such general discussion also yields the statement that the un-vanishing tree NLSM amplitudes only have the even number of external legs. Then, by imposing the universality of single soft behavior observed from the $4$-point case, we will construct the general ${\cal C}(\sigma'_{n-2},k_i)$ for $n\geqq 6$. The whole process only uses the mass dimension, the universality of soft behavior, as well as the permutation invariance among external scalars.

\subsection{$4$-point NLSM amplitude }
\label{subsec-34p}

For the $3$-point case, the numerator has mass dimension $2$. However, one can never use three on-shell massless
momenta satisfying momentum conservation to construct any un-vanishing Lorentz invariant with mass dimension $2$. Thus, the $3$-point NLSM amplitude does not exist.

The simplest NLSM amplitude is the $4$-point one ${\cal A}_{\rm N}(\sigma_4)$. It can be expanded to KK BAS basis, namely,
\bea
{\cal A}_{\rm N}(\sigma_4)=C_1\,{\cal A}_{\rm S}(1,2,3,4|\sigma_4)+C_2\,{\cal A}_{\rm S}(1,3,2,4|\sigma_4)\,.~~~~\label{expan-nlsm-4p}
\eea
The coefficients $C_1$ and $C_2$ have mass dimension $4$. The $4$-point case is special,
since one can choose a particular external leg $i$, then use on-shell and momentum conservation conditions to reduce any $k_a\cdot k_b$ to $k_i\cdot k_j$, with $a,b,j\in\{1,2,3,4\}$.
For instance, suppose we choose $i=3$, then we have
\bea
& &k_1\cdot k_2=k_3\cdot k_4\,,~~~~k_1\cdot k_4=k_3\cdot k_2\,,~~~~k_2\cdot k_4=k_3\cdot k_1\,.
\eea
Therefore, when considering the soft behavior for $k_i\to\tau k_i$, one can always express $C_1$ as
\bea
C_1=\sum_{a,b\in\{1,2,3,4\}\setminus i}\,\a_{ab}\,(k_i\cdot {\cal K}_a)\,(k_i\cdot{\cal K}_b)\,,~~~\label{c1}
\eea
where ${\cal K}_a$ and ${\cal K}_b$ are combinations of external momenta, $\a_{ab}$ has mass dimension $0$ and depend on subscripts $a$ and $b$. Thus the coefficient $C_1$ is obviously at the $\tau^2$ order. On the other hand, the leading soft behavior
of ${\cal A}_{\rm S}(1,2,3,4|\sigma_4)$ is at the $\tau^{-1}$ order. Combining $C_1$ and ${\cal A}_{\rm S}(1,2,3,4|\sigma_4)$ together, we see the soft behavior of the first term
at the r.h.s of \eref{expan-nlsm-4p} starts at the $\tau^{1}$ order. The consideration for the second term at the r.h.s of \eref{expan-nlsm-4p} is analogous. Thus, we conclude that the single soft behavior of $4$-point NLSM amplitude does not exist at $\tau^{-1}$ and $\tau^0$ orders. In other words,
in the limit $\tau\to0$, the $4$-point NLSM amplitude vanishes. Such behavior is known as Adler's zero.

The vanishing of the $4$-point amplitude in the single soft limit is another evidence for the vanishing of $3$-point one. Suppose the $n$-point and $(n+1)$-point NLSM amplitudes exist, by taking an external leg to be soft, one can always factorize the later one as the product of the former one and an un-vanishing leading soft factor. Thus, the vanishing of $(n+1)$-point amplitude in the single soft limit indicates the vanishing of $n$-point amplitude.

In this note, one of basic assumptions is the universality of soft behaviors. As the consequence of universality, the vanishing of single soft behavior at $\tau^{-1}$ and $\tau^0$ orders holds for any NLSM amplitude with arbitrary number of external legs. If we assume the existence of the $4$-point amplitude, then the $5$-point one must vanish, otherwise the vanishing of $5$-point amplitude in the single soft limit will imply the vanishing of $4$-point one. One can generalize the above argument further and conclude that the number of external legs for each un-vanishing NLSM amplitude should be even.

For now, we can not give the explicit formula of ${\cal C}(\sigma'_{2},k_i)$ for the $4$-point case. In subsection.\ref{subsec-np}, we will use the universality of soft behaviors to determine the $n$-point NLSM amplitudes in the expanded formula, with $n\geq6$. In the next section, we will use such expanded formula to derive the double soft factors for NLSM amplitudes at $\tau^0$ and $\tau^1$ orders. Imposing the universality of double soft factor, the $4$-point amplitude can be generated by removing two soft legs from $6$-point one, then ${\cal C}(\sigma'_{2},k_i)$ for the $4$-point case can be fixed through such manipulation, as will be shown in subsection.\ref{subsec-0order} in section.\ref{sec-double-soft}.

\subsection{$n$-point case}
\label{subsec-np}

As pointed out at the end of subsection.\ref{subsec-34p}, the universality of soft behaviors indicates that each NLSM amplitude vanishes at $\tau^{-1}$
and $\tau^0$ orders when one of external legs being soft. This subsection aims to find ${\cal C}(\sigma'_{n-2},k_i)$ for the $n$-point NLSM amplitude with $n\geq6$, by imposing the above
requirement, as well as the permutation symmetry among external legs in $\{2,\cdots,n-1\}$ which is evident in \eref{exp-N-KK}.

Coefficients ${\cal C}(\sigma'_{n-2},k_i)$ have mass dimension $2(n-2)$ and depend on color orderings $(1,\sigma'_{n-2},n)$ carried by ${\cal A}_{\rm S}(1,\sigma'_{n-2},n|\sigma_n)$. Consider $k_2\to\tau k_2$, and expand ${\cal A}_{\rm N}(\sigma_n)$ in $\tau$. In order to construct ${\cal A}_{\rm N}(\sigma_n)$ which
vanishes at $\tau^{-1}$ and $\tau^0$ orders, the simplest idea is to require each term in ${\cal C}(\sigma'_{n-2},k_i)$ to contain the factor
$(k_2\cdot {\cal K}_{a_2})\,(k_2\cdot{\cal K}_{b_2})$, similar as in \eref{c1} for the $4$-point case.
Then, situations $k_i\to\tau k_i$ for other $i\in\{2,\cdots,n-1\}$ yield the conclusion that each term in ${\cal C}(\sigma'_{n-2},k_i)$ also contains
$(k_i\cdot {\cal K}_{a_i})\,(k_i\cdot{\cal K}_{b_i})$. The total number of $k_i$ required by the above construction
is $2(n-2)$, which satisfies the correct mass dimension of ${\cal C}(\sigma'_{n-2},k_i)$. Then, each ${\cal C}(\sigma'_{n-2},k_i)$ can be decomposed as the linear combination of building blocks $\prod_{j=1}^{n-2}\,k_{a_j}\cdot k_{b_j}$, where the set $\{a_j,b_j\}$ contains two $i$ for each $i\in\{2,\cdots,n-1\}$, since when expanding to KK basis the coefficients do not contain any pole. When taking $k_1\to\tau k_1$ or $k_n\to\tau k_n$ and expanding in $\tau$, the coefficients ${\cal C}(\sigma'_{n-2},k_i)$ constructed in the above way are at the $\tau^0$ order, which makes the single soft behavior of ${\cal A}_{\rm N}(\sigma_n)$ to be the $\tau^{-1}$ order, therefore violates the universality of the soft behavior. Consequently, the above naive construction is not correct. Notice that
for $n\geq6$ cases one can not use on-shell and momentum conservation conditions to reduce the general $k_a\cdot k_b$
to $k_j\cdot\W {\cal K}$, as what we did for the $n=4$ case in subsection.\ref{subsec-34p}, where $a,b,j\in\{1,\cdots,n\}$ and $\W {\cal K}$ is a combination of external momenta.

To find the correct answer, we consider $k_2\to\tau k_2$ and express the leading order contribution of ${\cal A}_{\rm N}(\sigma_n)$ as
\bea
{\cal A}_{\rm N}^{(0)}(\sigma_n)=\sum_{\sigma'_{n-2}}\,{\cal C}^{(0)}(\sigma'_{n-2},k_i)\,S^{(0)}_{\rm S}(2)\,{\cal A}_{\rm S}(1,\sigma'_{n-2}\setminus2,n|\sigma_n\setminus2)\,,
\eea
where the soft theorem \eref{soft-theo-s} for external BAS scalar has been used.
The leading order contribution ${\cal C}^{(0)}(\sigma'_{n-2},k_i)$ is obtained from ${\cal C}(\sigma'_{n-2},k_i)$ as follows.
For each ${\cal K}_a\cdot{\cal K}_b$ contained in ${\cal C}(\sigma'_{n-2},k_i)$, where ${\cal K}_a$ and ${\cal K}_b$
are again two combinations of external momenta, we turn it to ${\cal K}_a\cdot{\cal K}_b\to{\cal K}^{(0)}_a\cdot{\cal K}^{(0)}_b$.
The combinatory momentum ${\cal K}^{(0)}_a$ is defined as
\bea
{\cal K}^{(0)}_a=\begin{cases}\displaystyle ~\tau k_2~~~~ &{\rm if}~{\cal K}_a=k_2 \,,\\
\displaystyle ~{\cal K}_a-k_2~~~~~ & {\rm otherwise}\,.\end{cases}~~~\label{defin-calk}
\eea
The definition of ${\cal K}^{(0)}_b$ is analogous.
The universality of soft behavior requires ${\cal A}_{\rm N}^{(0)}(1,\cdots,n)$ to be the $\tau^1$ order. The simplest and most natural solution to the above condition can be obtained by assuming the independence of BAS amplitudes in KK BAS basis. This assumption indicates that at $\tau^{-1}$ and $\tau^0$ orders, coefficients for different ${\cal A}_{\rm S}(1,\sigma'_{n-2}\setminus2,n|\sigma_n\setminus2)$
vanish individually, namely,
\bea
0&=&{\cal C}^{(0)}(2\shuffle\{\sigma_3,\cdots,\sigma_{n-1}\},k_i)\,S^{(0)}_{\rm S}(2)\,{\cal A}_{\rm S}(1,{\not 2}\shuffle\{\sigma_3,\cdots,\sigma_{n-1}\},n|\sigma_n\setminus2)\nn
&=&C(1,{\not 2})\,{1\over\tau}\,\Big({\delta_{12}\over s_{12}}+{\delta_{2\sigma_3}\over s_{2\sigma_3}}\Big)\,{\cal A}_{\rm S}(1,{\not 2},\sigma_3,\cdots,\sigma_{n-1},n|\sigma_n\setminus2)\nn
& &+\sum_{j=3}^{n-2}\,C(\sigma_j,{\not 2})\,{1\over\tau}\,\Big({\delta_{\sigma_j2}\over s_{\sigma_j2}}+{\delta_{2\sigma_{j+1}}\over s_{2\sigma_{j+1}}}\Big)\,{\cal A}_{\rm S}(1,\sigma_3,\cdots,\sigma_j,{\not 2},\sigma_{j+1},\cdots,\sigma_{n-1},n|\sigma_n\setminus2)\nn
& &+C(\sigma_{n-1},{\not 2})\,{1\over\tau}\,\Big({\delta_{\sigma_{n-1}2}\over s_{\sigma_{n-1}2}}+{\delta_{2n}\over s_{2n}}\Big)\,{\cal A}_{\rm S}(1,\sigma_3,\cdots,\sigma_{n-1},{\not 2},n|\sigma_n\setminus2)\,,~~~\label{assumption}
\eea
therefore
\bea
0=C(1,{\not 2})\,\Big({\delta_{12}\over s_{12}}+{\delta_{2\sigma_3}\over s_{2\sigma_3}}\Big)+
C(\sigma_{n-1},{\not 2})\,\Big({\delta_{\sigma_{n-1}2}\over s_{\sigma_{n-1}2}}+{\delta_{2n}\over s_{2n}}\Big)
+\sum_{j=3}^{n-2}\,C(\sigma_j,{\not 2})\,\Big({\delta_{\sigma_j2}\over s_{\sigma_j2}}+{\delta_{2\sigma_{j+1}}\over s_{2\sigma_{j+1}}}\Big)\,.~~~~\label{eq-c}
\eea
Here we abbreviated ${\cal C}^{(0)}(2\shuffle\{\sigma_3,\cdots,\sigma_{n-1}\},k_i)$ as $C(a,{\not 2})$ where $(a,{\not 2})$ emphasizes
the position of external leg $2$ in the color ordering $(\cdots,a,2,\cdots)$. The symbol $\shuffle$ in the first line of \eref{assumption} means the summation over particular permutations which preserves the ordering $(\sigma_3,\cdots,\sigma_{n-1})$, as explained below \eref{KK}. The notation ${\not2}$ means delating the leg $2$, thus all ${\cal A}_{\rm S}(1,{\not 2}\shuffle\{\sigma_3,\cdots,\sigma_{n-1}\},n|\sigma_n\setminus2)$ are the same BAS amplitude. Notice that since our first attempt is failed, we must assume
$C(a,{\not 2})$ are at the $\tau^0$ or $\tau^1$ order.

Since the ordering $\sigma_n$ is arbitrary, which means the values of $\delta_{ab}$ have not been fixed, to obtain the un-zero solution of $C(a,{\not 2})$ to the equation \eref{eq-c}, the only way is to employ the identity
\bea
\sum_{i\neq2}\,\delta_{i2}=0\,,
\eea
which is the special case of identity \eref{iden}.
In other words, the equation \eref{eq-c} should be reduced to
\bea
0=\W C\,\sum_{i\neq2}\,\delta_{i2}\,,~~~~\label{redu-eq-c}
\eea
where $\W C$ is a Lorentz invariant constructed from external momenta.
The formula \eref{redu-eq-c} requires that all $\delta_{i2}$ have a common coefficient $\W C$. Applying this condition to $\delta_{12}$, we immediately find $C(1,{\not 2})=2(k_2\cdot k_1)\,\W C$.
Then, applying the same condition to $\delta_{\sigma_32}$, we have
\bea
\W C\,={C(\sigma_3,{\not2})-C(1,{\not2})\over s_{2\sigma_3}}={C(\sigma_3,{\not2})-2\,(k_2\cdot k_1)\,\W C\over s_{2\sigma_3}}\,,
\eea
and the only solution is $C(\sigma_3,{\not2})=2(k_2\cdot K_{1\sigma_3})\,\W C$, where $K_{a_1\cdots a_m}\equiv\sum_{i=1}^m k_{a_i}$. Repeating the above process recursively, we get
\bea
C(a,{\not2})=2(k_2\cdot K_{1\sigma_3\cdots \sigma_a})\,\W C\,,~~~~\label{solu-c}
\eea
for arbitrary $a\in\{3,\cdots,n-1\}$.

Some remarks are in order. First, when expanding the NLSM amplitude to KK basis, coefficients do not contain any pole.
Thus, $\W C$ is a Lorentz invariant without any pole, otherwise one can not ensure that non of $C(a,{\not2})$ contains any pole. Secondly, we have assumed that $C(a,{\not 2})$ are at the $\tau^0$ or $\tau^1$
order previously. Since $C(a,{\not 2})$ includes the factor $k_2\cdot K_{1\sigma_3\cdots \sigma_a}$ accompanied with $\tau$ under the re-scaling $k_2\to\tau k_2$, we now exclude the $\tau^0$ case and conclude that $C(a,{\not 2})$ and $\W C$ are at
the $\tau^1$ order. Finally, for a given color ordering $(1,\sigma_3,\cdots,\sigma_{n-1},n)$ carried by ${\cal A}_{\rm S}(1,\sigma_3,\cdots,\sigma_{n-1},n|\sigma_n\setminus2)$, $\W C$ is independent of the position of leg $2$ in original orderings $(1,2\shuffle\{\sigma_3,\cdots,\sigma_{n-1}\},n)$.

The solution \eref{solu-c} indicates that coefficients ${\cal C}(\sigma'_{n-2},k_i)$ contain the component $k_2\cdot X_2$, where the combinatory momentum $X_i$ is defined as the summation of momenta carried by external legs at the l.h.s of $i$ in the color ordering.
This result can be generalized as ${\cal C}(\sigma'_{n-2},k_i)\propto k_i\cdot X_i$ for each $i\in\{2,\cdots,n-1\}$,
because of the permutation symmetry among external legs $i\in\{2,\cdots,n-1\}$ in the expansion \eref{exp-N-KK}. Since coefficients ${\cal C}(\sigma'_{n-2},k_i)$ have mass dimension $2(n-2)$, the above information fixes them completely, namely,
\bea
{\cal C}(\sigma'_{n-2},k_i)=\a\,\prod_{i=2}^{n-1}\,k_i\cdot X_i\,,~~~\label{solu-coe}
\eea
where $\a$ is an constant with mass dimension $0$. Since ${\cal C}(\sigma'_{n-2},k_i)$ does not contain any pole, $\a$ is independent of any kinematical variable. Furthermore, $\a$ is also independent of the color ordering $(1,\sigma'_{n-2},n)$, due to the permutation invariance.
Substituting the solution \eref{solu-coe} into \eref{exp-N-KK}, the expanded formula of $n$-point NLSM amplitude is found to be
\bea
{\cal A}_{\rm N}(\sigma_n)=\sum_{\sigma'_{n-2}}\,\Big(\prod_{i=2}^{n-1}\,k_i\cdot X_i\Big)\,{\cal A}_{\rm S}(1,\sigma'_{n-2},n|\sigma_n)\,.~~~~\label{expan-nlsm-np-resul}
\eea
Here we have fixed the constant $\a$ as $\a=1$, via an overall re-scaling of the amplitude. The expression in \eref{expan-nlsm-np-resul}
is the desired expanded formula of NLSM amplitude with $n\geq6$. In subsection.\ref{subsec-0order}, we will show that \eref{expan-nlsm-np-resul}
is also correct for the $4$-point case.

It is not hard to verify that ${\cal A}_{\rm N}(\sigma_n)$ given in \eref{expan-nlsm-np-resul} vanishes
at $\tau^{-1}$ and $\tau^0$ orders under the re-scaling $k_1\to\tau k_1$ or $k_n\to\tau k_n$, thus the expanded formula \eref{expan-nlsm-np-resul}
satisfies the correct single soft behavior for any external leg being soft. Taking $k_2\to\tau k_2$, $\W C$ in the solution \eref{solu-c} can be calculated from\eref{expan-nlsm-np-resul} as
\bea
\W C={\tau\over2}\,\prod_{i=3}^{n-1}\,k_i\cdot K_{1\sigma_3\cdots\sigma_{i-1}}\,,
\eea
which satisfies our expectations: $\W C$ is a Lorentz invariant without any pole and is independent of the position of inserting the leg $2$. The situations for other $k_i\to\tau k_i$ are analogous.

The solution \eref{solu-coe} is obtained by assuming the independence of BAS amplitudes in KK BAS basis. Indeed, such independence is violated by the well known Bern-Carrasco-Johansson (BCJ) relations among BAS amplitudes. BCJ relations allow more solutions to equation \eref{eq-c}. For instance, one can modify the expanded formula in \eref{expan-nlsm-np-resul} by adding terms which vanish automatically,
such as
\bea
{\cal A}_{\rm N}(\sigma_n)&=&\sum_{\sigma'_{n-2}}\,\Big(\prod_{i=2}^{n-1}\,k_i\cdot X_i\Big)\,{\cal A}_{\rm S}(1,\sigma'_{n-2},n|\sigma_n)\nn
& &+(k_2\cdot X_2)\,{\cal T}\,{\cal A}_{\rm S}(1,\sigma'_{n-2},n|\sigma_n)\,,~~~\label{modify-expand}
\eea
where ${\cal T}$ is a Lorentz invariant with mass dimension $2(n-3)$. The above modification is guaranteed by the fundamental BCJ relation
\bea
0=(k_2\cdot X_2)\,{\cal A}_{\rm S}(1,2\shuffle\{\sigma_3,\cdots,\sigma_{n-1}\},n|\sigma_n)\,.
\eea
Thus, when considering BCJ relations, the expanded formula \eref{expan-nlsm-np-resul} should be understood as an equivalent class rather than an unique expression.

The disadvantage of the modified formula \eref{modify-expand} is the breaking of manifest permutation symmetry among legs in $\{2,\cdots,n-1\}$.
Let us only focus on the expanded formulas those manifest such permutation invariance, and ask if the BCJ relations lead to different solutions to equation \eref{eq-c} which satisfy this symmetry, and the coefficients ${\cal C}(\sigma'_{n-2},k_i)$ do not contain any pole. Quite surprisingly, for the $6$-point and $8$-point cases, such new solution which is un-equivalent to \eref{expan-nlsm-np-resul} can not be found. Thus, we conjecture that this situation is general, and claim that the expansion in \eref{expan-nlsm-np-resul} is the only solution which manifests the desired permutation invariance, although the general proof is absent.

\section{Double soft factors}
\label{sec-double-soft}

As discussed in subsection.\ref{subsec-34p} in the previous section, using the single soft theorem, one can generate the $n$-point tree amplitude from the $(n+1)$-point one, by removing the soft external leg. The tree NLSM amplitude do not have such luxuries, since the number of external legs must be even. However, it is natural to think the $n$-point NLSM amplitude as the resulting object of removing two soft legs from the $(n+2)$-point one. This picture leads to the consideration of the double soft behavior of NLSM tree amplitudes.

The double soft theorems for tree NLSM amplitudes have been obtained via different methods \cite{Cachazo:2015ksa,Du:2015esa}. The leading soft factor is at the $\tau^0$ order, while the sub-leading one is the $\tau^1$ order. Such leading order behavior is coincide with the picture of generating the $n$-point amplitude from the $(n+2)$-point one, since such manipulation forbids the vanishing of the $(n+2)$-point amplitude in the limit $\tau\to 0$. In this section, we propose an efficient new approach to derive the leading and sub-leading double soft factors, based on the expansion in \eref{expan-nlsm-np-resul}.

Without losing of generality, we consider $k_a\to\tau k_a,\,k_b\to\tau k_b$ and expand ${\cal A}_{\rm N}(1,\cdots,n,a,b)$ by $\tau$.
We chose the basis as BAS amplitudes whose external legs $1$ and $n$ are fixed at two ends in the first color ordering, thus the $n+2$-point amplitude ${\cal A}_{\rm N}(1,\cdots,n,a,b)$ can be expanded as
\bea
{\cal A}_{\rm N}(1,\cdots,n,a,b)=(k_a\cdot X_a)\,(k_b\cdot X_b)\,\Big(\prod_{i=2}^{n-1}\,k_i\cdot X_i\Big)\,
{\cal A}_{\rm S}(1,a\shuffle b\shuffle2\shuffle\cdots\shuffle n-1,n|1,\cdots,n,a,b)\,.~~~\label{expan-for-ds}
\eea
According to the meaning of $\shuffle$ mentioned below \eref{KK}, one can rewrite the expansion in \eref{expan-nlsm-np-resul} as
\bea
{\cal A}_{\rm N}(\sigma_n)=\Big(\prod_{i=2}^{n-1}\,k_i\cdot X_i\Big)\,{\cal A}_{\rm S}(1,2\shuffle\cdots\shuffle n-1,n|\sigma_n)\,.
\eea
We used this notation in \eref{expan-for-ds}, to emphasize the positions of special legs $a$ and $b$ in the first color ordering carried by ${\cal A}_{\rm S}(1,\sigma_1,\cdots,\sigma_n,n|1,\cdots,n,a,b)$.
For later convenience, in the reminder of this section we denote the BAS amplitude ${\cal A}_{\rm S}(1,\sigma_1,\cdots,\sigma_n,n|1,\cdots,n,a,b)$ as ${\rm A}(\sigma_1,\cdots,\sigma_n)$, where $\{\sigma_1,\cdots,\sigma_n\}=\{a,b,2,\cdots,n-1\}$.

\subsection{Leading order}
\label{subsec-0order}

%
\begin{figure}
  \centering
  \includegraphics[width=9cm]{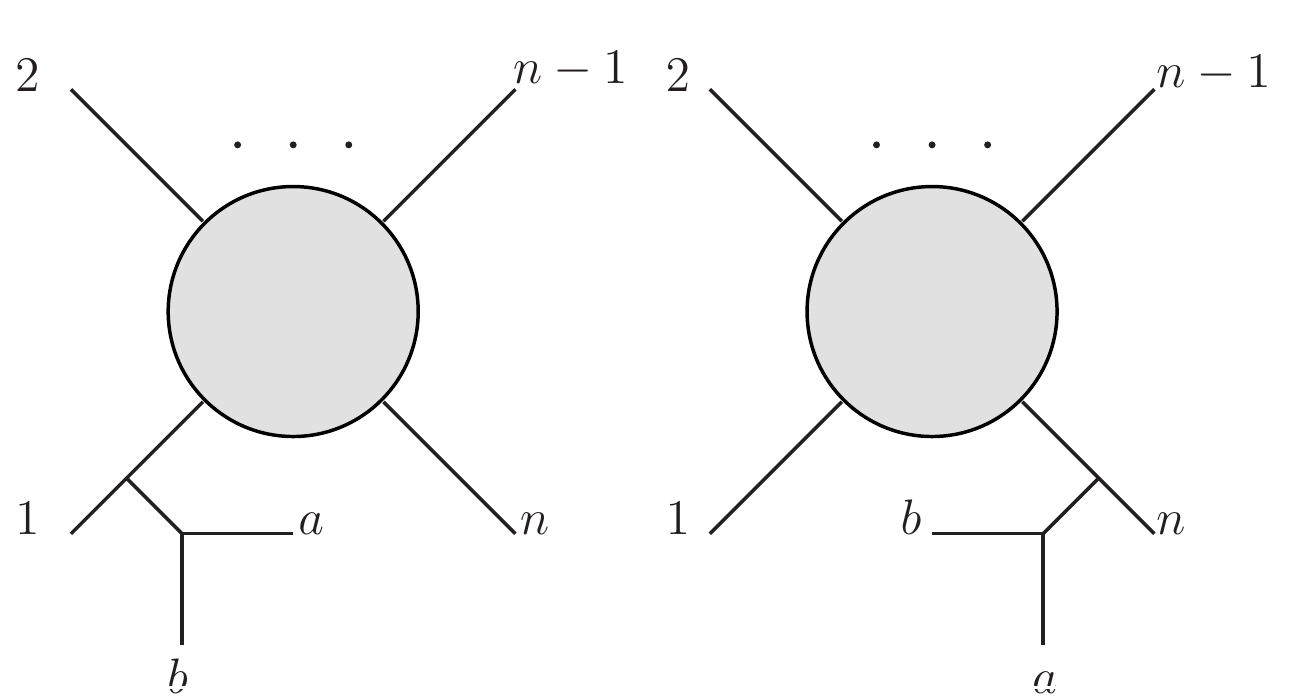} \\
  \caption{First type of Feynman diagrams.}\label{abY}
\end{figure}

To derive the double soft factor at leading order, let us consider Feynman diagrams for BAS amplitudes ${\rm A}(a\shuffle b\shuffle2\shuffle\cdots\shuffle n-1)$ those external legs
$a$ and $b$ are coupled to a common vertex. Such diagrams allowed by the second color ordering are shown in Figure.\ref{abY}. Among terms at the r.h.s of the expansion \eref{expan-for-ds}, contributions from
the first configuration in Figure.\ref{abY} are contained in
\bea
& &(k_a\cdot k_1)\,(k_b\cdot K_{1a})\,\Big(\prod_{i=2}^{n-1}\,k_i\cdot X_i\Big)\,{\rm A}(a,b,2\shuffle\cdots\shuffle n-1)\nn
&+&(k_b\cdot k_1)\,(k_a\cdot K_{1b})\,\Big(\prod_{i=2}^{n-1}\,k_i\cdot X_i\Big)\,{\rm A}(b,a,2\shuffle\cdots\shuffle n-1)\,,~~\label{0order-confi1}
\eea
while contributions from the second configuration are included in
\bea
& &(k_b\cdot k_n)\,(k_a\cdot K_{bn})\,\Big(\prod_{i=2}^{n-1}\,k_i\cdot X_i\Big)\,{\rm A}(2\shuffle\cdots\shuffle n-1,a,b)\nn
&+&(k_a\cdot k_n)\,(k_b\cdot K_{an})\,\Big(\prod_{i=2}^{n-1}\,k_i\cdot X_i\Big)\,{\rm A}(2\shuffle\cdots\shuffle n-1,b,a)\,.~~\label{0order-confi2}
\eea
After taking $k_a\to\tau k_a,\,k_b\to\tau k_b$ and expanding in $\tau$, the leading order contribution from \eref{0order-confi1}
can be calculated as
\bea
L_1&=&\tau^2\,(k_a\cdot k_1)\,\big(k_b\cdot (k_1+\tau k_a)\big)\,\Big(\prod_{i=2}^{n-1}\,k_i\cdot X^{(0)}_i\Big)\,{1\over s^{(0)}_{ab}}\,{1\over s^{(0)}_{ab1}}\,{\rm A}(2\shuffle\cdots\shuffle n-1)\nn
& &-\tau^2\,(k_b\cdot k_1)\,\big(k_a\cdot (k_1+\tau k_b)\big)\,\Big(\prod_{i=2}^{n-1}\,k_i\cdot X^{(0)}_i\Big)\,{1\over s^{(0)}_{ab}}\,{1\over s^{(0)}_{ab1}}\,{\rm A}(2\shuffle\cdots\shuffle n-1)\nn
&=&{(k_a-k_b)\cdot k_1\over 4(k_a+k_b)\cdot k_1}\,\Big[\Big(\prod_{i=2}^{n-1}\,k_i\cdot X^{(0)}_i\Big)\,{\rm A}(2\shuffle\cdots\shuffle n-1)\Big]\nn
&=&{(k_a-k_b)\cdot k_1\over 4(k_a+k_b)\cdot k_1}\,{\cal A}_{\rm N}(1,\cdots,n)\,,~~\label{L1}
\eea
where $X^{(0)}_i$ arise from $X_i$ by deleting $k_a$ and $k_b$, $s^{(0)}_{ab}$ and $s^{(0)}_{ab1}$ stand for leading order contributions of $s_{ab}$ and $s_{ab1}$ respectively therefore $s^{(0)}_{ab}=2\tau^2\,k_a\cdot k_b$, $s^{(0)}_{ab1}=2\tau\,(k_a+k_b)\cdot k_1$. Here the relative $-$ sign in the first equality can be figured out via the diagrammatic rules, and a more direct way to see it is employing the anti-symmetry of structure constant $f^{abc}$ of Lie group which indicates a $-$ sign when swamping legs $a$ and $b$. Because of this $-$ sign, terms at the $\tau^{-1}$ order cancel each other, leaving the un-vanishing $L_1$ at the $\tau^0$ order. The last equality uses the observation
\bea
{\cal A}_{\rm N}(1,\cdots,n)=\Big(\prod_{i=2}^{n-1}\,k_i\cdot X^{(0)}_{i}\Big)\,{\rm A}(2\shuffle\cdots\shuffle n-1)\,,~~\label{expan-nlsm-np1}
\eea
based on the expansion \eref{expan-nlsm-np-resul} and the definition of $X^{(0)}_i$.

The consideration for \eref{0order-confi2} is analogous and gives
\bea
L_2
&=&{(k_b-k_a)\cdot k_n\over 4(k_a+k_b)\cdot k_n}\,{\cal A}_{\rm N}(1,\cdots,n)\,,~~\label{L2}
\eea
which is also at the $\tau^0$ order. We claim that \eref{L1} and \eref{L2} provide the full double soft behavior of ${\cal A}_{\rm N}(1,\cdots,n,a,b)$
at leading order. The reason can be explained as follows. For diagrams those legs $a$ and $b$ are coupled to different vertices, the double soft limit can be achieved by taking single soft limits for $a$ and $b$ successively. Since the single soft behavior for $a$ or $b$ vanishes at the $\tau^0$ order, one can conclude that taking two single soft limits consecutively contributes nothing at the $\tau^0$ order. Thus, the leading double soft factor
can be found by combining \eref{L1} and \eref{L2} together, which gives
\bea
{\cal A}^{(0)}_{\rm N}(1,\cdots,n,a,b)=S^{(0)}_{\rm N}(a,b)\,{\cal A}_{\rm N}(1,\cdots,n)\,,~~\label{dsoft-0}
\eea
where
\bea
S^{(0)}_{\rm N}(a,b)={(k_a-k_b)\cdot k_1\over 4(k_a+k_b)\cdot k_1}+{(k_b-k_a)\cdot k_n\over 4(k_a+k_b)\cdot k_n}\,.~~\label{dsoft-op-0}
\eea
This result is the same as that obtained in \cite{Cachazo:2015ksa,Du:2015esa} via different approaches.

The universality of leading soft factor \eref{dsoft-op-0} indicates that the expanded formula \eref{expan-nlsm-np-resul} is valid for the $4$-point NLSM amplitude. The reason is as follows. The universality imposes
\bea
{\cal A}^{(0)}_{\rm N}(1,2,3,4,a,b)=S^{(0)}_{\rm N}(a,b)\,{\cal A}_{\rm N}(1,2,3,4)\,,~~\label{dsoft-0-4p}
\eea
which is a special case of \eref{dsoft-0}. Since the relation \eref{dsoft-0} is based on the expansion \eref{expan-nlsm-np1},
${\cal A}_{\rm N}(1,2,3,4)$ in \eref{dsoft-0-4p} must satisfy the expansion \eref{expan-nlsm-np1} which is equivalent to \eref{expan-nlsm-np-resul}.

\subsection{Sub-leading order}
\label{subsec-1order}

The sub-leading double soft factor is more complicate, since Feynman diagrams those $a$ and $b$ are coupled to different vertices also have un-vanishing contributions at the $\tau^1$ order. We will consider the corresponding diagrams one by one.

Let us start with the first configuration in Figure.\ref{abY}, and express corresponding terms contained in \eref{0order-confi1} as
\bea
& &\Big[(k_a\cdot k_1)\,(k_b\cdot K_{1a})\,\Big(\prod_{i=2}^{n-1}\,k_i\cdot X_i\Big)
-(k_b\cdot k_1)\,(k_a\cdot K_{1b})\,\Big(\prod_{i=2}^{n-1}\,k_i\cdot X_i\Big)\Big]\,{1\over s_{ab}}\,{1\over s_{ab1}}\,{\cal M}\nn
&=&{(k_a-k_b)\cdot k_1\over 2s_{ab1}}\,\Big(\prod_{i=2}^{n-1}\,k_i\cdot X_i\Big)\,{\cal M}\,.
~~\label{1-order-confi1}
\eea
The relative $-$ sign was interpreted in subsection.\ref{subsec-0order} around \eref{L1}. When $k_a\to\tau k_a,\,k_b\to\tau k_b$, \eref{1-order-confi1} behaves as
\bea
{(k_a-k_b)\cdot k_1\over 4(k_a+k_b)\cdot k_1+4\tau\,k_a\cdot k_b}\,\Big(\prod_{i=2}^{n-1}\,k_i\cdot X_i(\tau)\Big)\,{\cal M}(\tau)~~\label{1-order-confi11}
\eea
Our purpose is to extract the $\tau^1$
order terms from \eref{1-order-confi11}. First, one can expand ${\cal M}(\tau)$
as
\bea
{\cal M}(\tau)={\cal M}(0)+\tau\,{\partial\over\partial\tau}\,{\cal M}(\tau)\big|_{\tau=0}+\cdots\,
\eea
and pick up the second term. One can observe that the parameter $\tau$ enters ${\cal M}(\tau)$ only through the combinatory
momentum $k_1+\tau(k_a+k_b)$, thus
\bea
{\partial\over\partial\tau}={1\over\tau}\,(k_a+k_b)\cdot{\partial\over\partial(k_a+k_b)}=(k_a+k_b)\cdot{\partial\over\partial k_1}\,,
\eea
which leads to
\bea
{\partial\over\partial\tau}\,{\cal M}(\tau)\big|_{\tau=0}=(k_a+k_b)\cdot{\partial\over\partial k_1}\,{\rm A}(2\shuffle\cdots\shuffle n-1)\,,
\eea
where the observation ${\cal M}(0)={\rm A}(2\shuffle\cdots\shuffle n-1)$ was used. Consequently, the first piece is found to be
\bea
{\cal P}_{1;1}=\tau\,{(k_a-k_b)\cdot k_1\over 4(k_a+k_b)\cdot k_1}\,\Big(\prod_{i=2}^{n-1}\,k_i\cdot X^{(0)}_i\Big)\,\Big[(k_a+k_b)\cdot\partial_{ k_1}\,{\rm A}(2\shuffle\cdots\shuffle n-1)\Big]\,,~~\label{p11}
\eea
where
\bea
\partial_{k_1}\equiv{\partial\over\partial k_1}\,.
\eea
Secondly, one can expand the denominator in \eref{1-order-confi11} and pick up the second terms at the r.h.s of
\bea
{1\over 4(k_a+k_b)\cdot k_1+4\tau\,k_a\cdot k_b}={1\over 4(k_a+k_b)\cdot k_1}-{\tau\,k_a\cdot k_b\over4\big((k_a+k_b)\cdot k_1\big)^2}+\cdots\,,
\eea
which gives the second piece
\bea
{\cal P}_{2;1}=-\tau\,{\big((k_a-k_b)\cdot k_1\big)\,(k_a\cdot k_b)\over 4\big((k_a+k_b)\cdot k_1\big)^2}\,\Big(\prod_{i=2}^{n-1}\,k_i\cdot X^{(0)}_i\Big)\,{\rm A}(2\shuffle\cdots\shuffle n-1)\,.~~\label{p21}
\eea
Finally, each $X_i(\tau)$ contains $\tau\,(k_a+k_b)$, thus the third piece is found to be
\bea
{\cal P}_{3;1}&=&\tau\,{(k_a-k_b)\cdot k_1\over 4(k_a+k_b)\cdot k_1}\,\Big[\sum_{j=2}^{n-1}\,\Big(k_j\cdot(k_a+k_b)\,{\prod_{i=2}^{n-1}\,k_i\cdot X^{(0)}_i\over k_j\cdot X^{(0)}_j}\Big)\Big]\,{\rm A}(2\shuffle\cdots\shuffle n-1)\nn
&=&\tau\,{(k_a-k_b)\cdot k_1\over 4(k_a+k_b)\cdot k_1}\,\Big[(k_a+k_b)\cdot\partial_{k_1}\,\Big(\prod_{i=2}^{n-1}\,k_i\cdot X^{(0)}_i\Big)\Big]\,{\rm A}(2\shuffle\cdots\shuffle n-1)\,.~~\label{p31}
\eea

Similarly, considering the second configuration in Figure.\ref{abY} yields
\bea
{\cal P}_{1;2}=\tau\,{(k_b-k_a)\cdot k_n\over 4(k_a+k_b)\cdot k_n}\,\Big(\prod_{i=2}^{n-1}\,k_i\cdot X^{(0)}_i\Big)\,\Big[(k_a+k_b)\cdot\partial_{k_n}\,{\rm A}(2\shuffle\cdots\shuffle n-1)\Big]\,,~~\label{p12}
\eea
\bea
{\cal P}_{2;2}=-\tau\,{\big((k_b-k_a)\cdot k_n\big)\,(k_a\cdot k_b)\over 4\big((k_a+k_b)\cdot k_n\big)^2}\,\Big(\prod_{i=2}^{n-1}\,k_i\cdot X^{(0)}_i\Big)\,{\rm A}(2\shuffle\cdots\shuffle n-1)\,,~~\label{p22}
\eea
as well as
\bea
{\cal P}_{3;2}&=&\tau\,{(k_b-k_a)\cdot k_n\over 4(k_a+k_b)\cdot k_n}\,\Big[\sum_{j=2}^{n-1}\,\Big(k_j\cdot(k_a+k_b)\,{\prod_{i=2}^{n-1}\,k_i\cdot X^{(0)}_i\over k_j\cdot X^{(0)}_j}\Big)\Big]\,{\rm A}(2\shuffle\cdots\shuffle n-1)\nn
&=&\tau\,{(k_b-k_a)\cdot k_n\over 4(k_a+k_b)\cdot k_n}\,\Big[(k_a+k_b)\cdot\partial_{k_n}\,\Big(\prod_{i=2}^{n-1}\,k_i\cdot X^{(0)}_i\Big)\Big]\,{\rm A}(2\shuffle\cdots\shuffle n-1)
\,.~~\label{p32}
\eea
The process is paralleled to those for obtaining \eref{p11}, \eref{p21} and \eref{p31}. To derive \eref{p32}, one should use momentum conservation to replace the explicit form of $X_i$ by the equivalent one ${\cal X}_i=X_i-\big(k_a+k_b+\sum_{i=1}^n\,k_i\big)$. The special expressions for $X_i$ imply that the individual results in
\eref{p31} and \eref{p32} are inconsistent with momentum conservation. In \eref{p32}, suppose one use momentum conservation to remove $k_n$ in $X^{(0)}_i$, then $k_i\cdot X^{(0)}_i$ will be annihilated by the operator $\partial_{k_n}$. Similar phenomenon happens for \eref{p31}. However, after combining \eref{p31} and \eref{p32} together, the resulting object is consistent with momentum conservation. One can use momentum conservation to modify the explicit expression of any $X_i$ freely, the combination ${\cal P}_{4;1}+{\cal P}_{4;2}$ always gives the correct result. Similarly, the individual pieces \eref{p11} and \eref{p12} are inconsistent with momentum conservation, while
the combination ${\cal P}_{1;1}+{\cal P}_{1;2}$ does not have this problem.

\begin{figure}
  \centering
  \includegraphics[width=9cm]{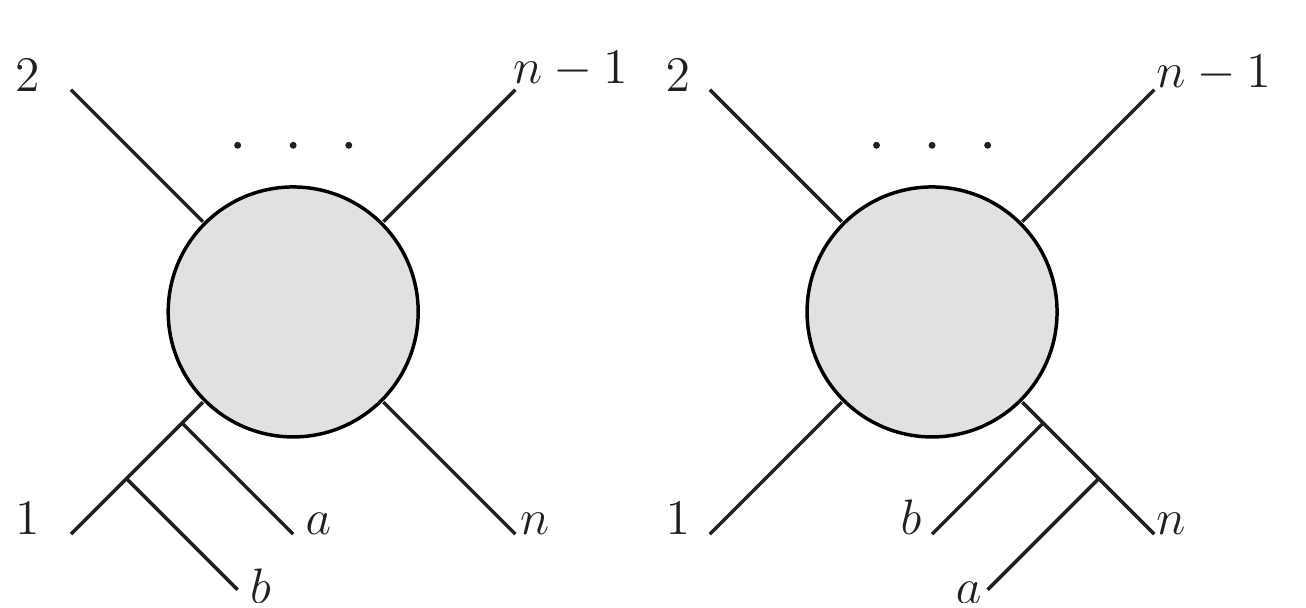} \\
  \caption{Second type of Feynman diagrams.}\label{abII}
\end{figure}

Now we turn to a new type of Feynman diagrams which can be seen in Figure.\ref{abII}. We focus on the first configuration in Figure.\ref{abII},
and the second one can be treated similarly. At the r.h.s of \eref{expan-for-ds}, contributions from such diagrams are included in
\bea
(k_b\cdot k_1)\,(k_a\cdot K_{1b})\,\Big(\prod_{i=2}^{n-1}\,k_i\cdot X_i\Big)\,{\rm A}(b,a,2\shuffle\cdots\shuffle n-1)\,,
\eea
thus can be expressed as
\bea
(k_b\cdot k_1)\,(k_a\cdot K_{1b})\,\Big(\prod_{i=2}^{n-1}\,k_i\cdot X_i\Big)\,{1\over s_{b1}}\,{1\over s_{ab1}}\,{\cal N}
&=&{k_a\cdot K_{1b}\over 2s_{ab1}}\,\Big(\prod_{i=2}^{n-1}\,k_i\cdot X_i\Big)\,{\cal N}\,.~~\label{1-order-confi2}
\eea
Taking $k_a\to\tau k_a,\,k_b\to\tau k_b$ turns \eref{1-order-confi2} to
\bea
{k_a\cdot k_1+\tau\,k_a\cdot k_b\over 4(k_a+k_b)\cdot k_1+4\tau\,k_a\cdot k_b}\,\Big(\prod_{i=2}^{n-1}\,k_i\cdot X_i(\tau)\Big)\,{\cal N}(\tau)\,,~~\label{1-order-confi21}
\eea
and one can extract four pieces at the $\tau^1$ order. The first one, expanding ${\cal N}(\tau)$ by $\tau$ gives
\bea
{\cal P}_{4;1}=-\tau\,{k_a\cdot k_1\over 4(k_a+k_b)\cdot k_1}\,\Big(\prod_{i=2}^{n-1}\,k_i\cdot X^{(0)}_i\Big)\,\Big[(k_a+k_b)\cdot\partial_{k_1}\,{\rm A}(2\shuffle\cdots\shuffle n-1)\Big]\,,~~\label{p41}
\eea
where the observation ${\cal N}(0)=-{\rm A}(2\shuffle\cdots\shuffle n-1)$ is used, and the $-$ sign can be verified via diagrammatical rules. The second piece arises from
expanding the denominator of \eref{1-order-confi21}, which is given by
\bea
{\cal P}_{5;1}=-\tau\,{k_a\cdot k_b\over 4(k_a+k_b)\cdot k_1}\,\Big(\prod_{i=2}^{n-1}\,k_i\cdot X^{(0)}_i\Big)\,{\rm A}(2\shuffle\cdots\shuffle n-1)\,.~~\label{p51}
\eea
The third one is obtained by expanding the denominator of \eref{1-order-confi21}, which is found to be
\bea
{\cal P}_{6;1}=\tau\,{(k_a\cdot k_1)\,(k_a\cdot k_b)\over 4\big((k_a+k_b)\cdot k_1\big)^2}\Big(\prod_{i=2}^{n-1}\,k_i\cdot X^{(0)}_i\Big)\,{\rm A}(2\shuffle\cdots\shuffle n-1)\,.~~\label{p61}
\eea
The final one comes from $\tau\,(k_a+k_b)$ contained in $X_{i}(\tau)$ thus is given as
\bea
{\cal P}_{7;1}=-\tau\,{k_a\cdot k_1\over 4(k_a+k_b)\cdot k_1}\,\Big[(k_a+k_b)\cdot\partial_{k_1}\,\Big(\prod_{i=2}^{n-1}\,k_i\cdot X^{(0)}_i\Big)\Big]\,{\rm A}(2\shuffle\cdots\shuffle n-1)\,.~~\label{p71}
\eea
Considering the second configuration in Figure.\ref{abII} gives analogous results
\bea
{\cal P}_{4;2}=-\tau\,{k_b\cdot k_n\over 4(k_a+k_b)\cdot k_n}\,\Big(\prod_{i=2}^{n-1}\,k_i\cdot X^{(0)}_i\Big)\,\Big[(k_a+k_b)\cdot\partial_{k_n}\,{\rm A}(2\shuffle\cdots\shuffle n-1)\Big]\,,~~\label{p42}
\eea
\bea
{\cal P}_{5;2}=-\tau\,{k_a\cdot k_b\over 4(k_a+k_b)\cdot k_n}\,\Big(\prod_{i=2}^{n-1}\,k_i\cdot X^{(0)}_i\Big)\,{\rm A}(2\shuffle\cdots\shuffle n-1)\,,~~\label{p52}
\eea
\bea
{\cal P}_{6;2}=\tau\,{(k_b\cdot k_n)\,(k_a\cdot k_b)\over 4\big((k_a+k_b)\cdot k_n\big)^2}\Big(\prod_{i=2}^{n-1}\,k_i\cdot X^{(0)}_i\Big)\,{\rm A}(2\shuffle\cdots\shuffle n-1)\,,~~\label{p62}
\eea
and
\bea
{\cal P}_{7;2}=-\tau\,{k_b\cdot k_n\over 4(k_a+k_b)\cdot k_n}\,\Big[(k_a+k_b)\cdot\partial_{k_n}\,\Big(\prod_{i=2}^{n-1}\,k_i\cdot X^{(0)}_i\Big)\Big]\,{\rm A}(2\shuffle\cdots\shuffle n-1)\,.~~\label{p72}
\eea
Combinations ${\cal P}_{4;1}+{\cal P}_{4;2}$ and ${\cal P}_{7;1}+{\cal P}_{7;2}$ allow the expressions of
$X^{(0)}_i$ to be rewritten via momentum conservation, while individual pieces do not.

\begin{figure}
  \centering
  \includegraphics[width=5cm]{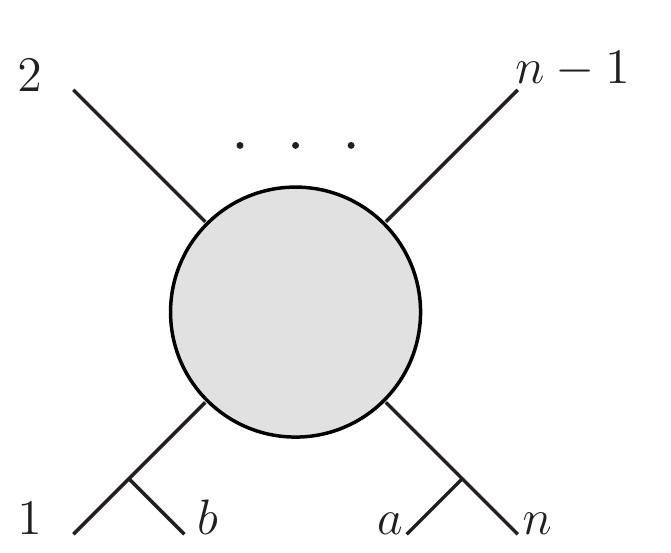} \\
  \caption{Third type of Feynman diagrams.}\label{anb1}
\end{figure}

Then we consider the configuration in Figure.\ref{anb1}, which corresponds to
\bea
-(k_b\cdot k_1)\,(k_a\cdot k_n)\,\Big(\prod_{i=2}^{n-1}\,k_i\cdot X_i\Big)\,{\rm A}(b,2\shuffle\cdots\shuffle n-1,a)\,,
\eea
at the r.h.s of \eref{expan-for-ds}. Here we have used on-shell and momentum conservation conditions to rewrite $X_a$ as $-k_n$. Under the re-scaling $k_a\to\tau k_a,\,k_b\to\tau k_b$, the leading order contributions from these terms are the $\tau^1$
order, provides
\bea
{\cal P}_8={\tau\over4}\,\Big[\Big(k_b\cdot\partial_{k_1}+k_a\cdot\partial_{k_n}\Big)\,\Big(\prod_{i=2}^{n-1}\,k_i\cdot X^{(0)}_i\Big)\Big]\,{\rm A}(2\shuffle\cdots\shuffle n-1)\,.~~\label{p8}
\eea
Only one of operators $\partial_{k_1}$ and $\partial_{k_n}$ is effective, since the definition of $X^{(0)}_i$ implies that $k_1$ and $k_n$
can not be included in $X^{(0)}_i$ simultaneously. The formula \eref{p8} gives
correct result for any expression of $X^{(0)}_i$, thus is consistent with momentum conservation.

\begin{figure}
  \centering
  \includegraphics[width=9cm]{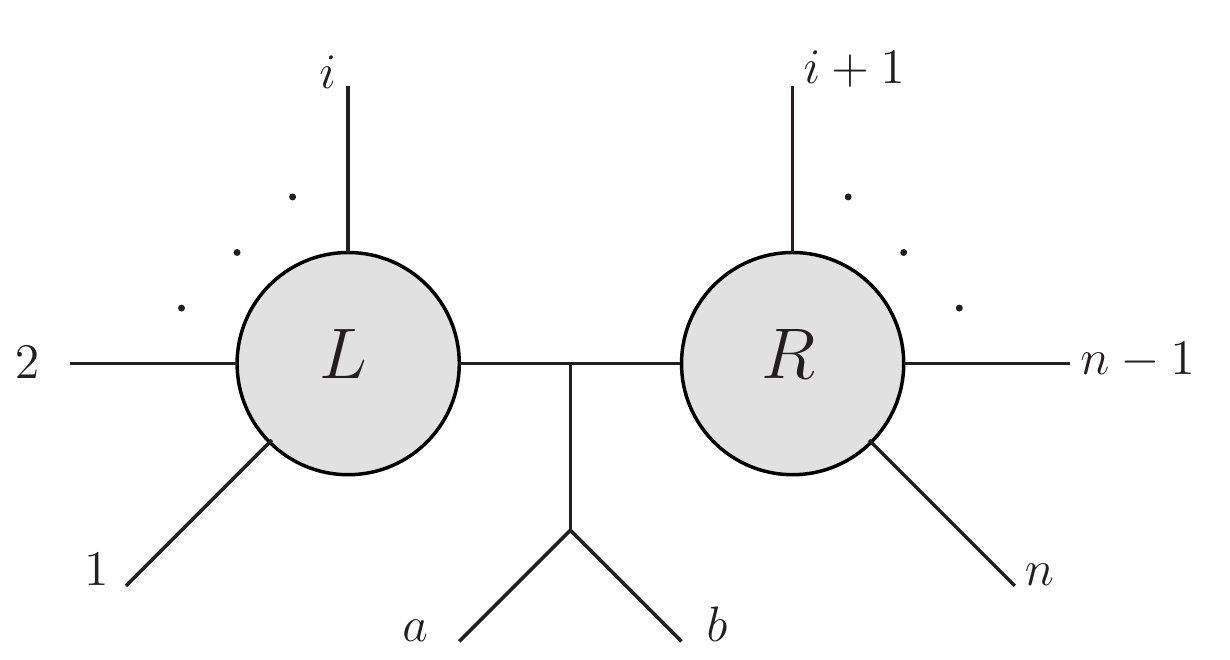} \\
  \caption{Fourth type of Feynman diagrams.}\label{abYm}
\end{figure}

\begin{figure}
  \centering
  \includegraphics[width=6cm]{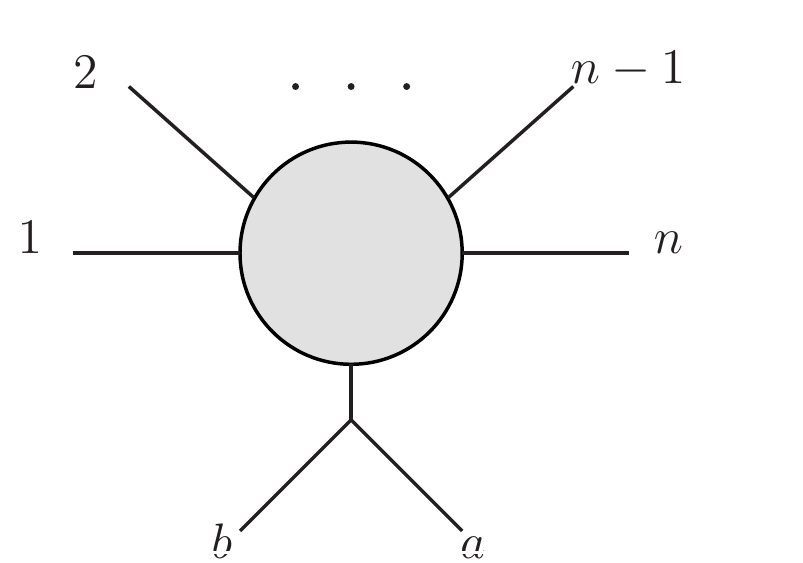} \\
  \caption{Feynman diagrams which should be excluded.}\label{eli}
\end{figure}

Next, we deal with the configuration in Figure.\ref{abYm} in which legs $a$ and $b$ coupled to a common vertex then coupled to an internal line. At the r.h.s of \eref{expan-for-ds}, this type of Feynman diagrams correspond to
\bea
& &(k_a\cdot K_{1\cdots i})\,(k_b\cdot K_{1\cdots ia})\,\Big(\prod_{j=2}^{n-1}\,k_j\cdot X_j\Big)\,{\rm A}(2\shuffle\cdots\shuffle i,a,b,i+1\shuffle\cdots\shuffle n-1)\nn
&+&(k_b\cdot K_{1\cdots i})\,(k_a\cdot K_{1\cdots ib})\,\Big(\prod_{j=2}^{n-1}\,k_j\cdot X_j\Big)\,{\rm A}(2\shuffle\cdots\shuffle i,b,a,i+1\shuffle\cdots\shuffle n-1)\,.
\eea
When $k_a\to\tau k_a,\,k_b\to\tau k_b$, the leading order contributions correspond to Figure.\ref{abYm} are at the $\tau^1$ order, and can be calculated as
\bea
& &\tau\,{(k_a-k_b)\cdot K_{1\cdots i}\over 2s_{1\cdots i}^2}\,\Big(\prod_{j=2}^{n-1}\,k_j\cdot X^{(0)}_j\Big)\,{\cal M}_L\,{\cal M}_R\nn
&=&-{\tau\over4}\,\Big((k_a-k_b)\cdot\partial_{k_1}\,{1\over s_{1\cdots i}}\Big)\Big(\prod_{j=2}^{n-1}\,k_j\cdot X^{(0)}_j\Big)\,{\cal M}_L\,{\cal M}_R\nn
&=&-{\tau\over4}\,\Big((k_b-k_a)\cdot\partial_{k_n}\,{1\over s_{i+1\cdots n}}\Big)\Big(\prod_{j=2}^{n-1}\,k_j\cdot X^{(0)}_j\Big)\,{\cal M}_L\,{\cal M}_R\,.
\eea
The last equality is obtained by replacing $K_{1\cdots i}$ with $-K_{i+1\cdots n}$ via momentum conservation. Notice that since
$s_{1\cdots i}$ is defined as $s_{1\cdots i}=K_{1\cdots i}^2$ which contains $k_1^2$, we have
$\partial_{k_{1,\mu}}\,s_{1\cdots i}=2K^\mu_{1\cdots i}$,
although $k_1^2=0$ and we usually omit it when expressing $s_{1\cdots i}$ explicitly. The analogous situation holds for $\partial_{k_{n,\mu}}\,s_{i+1\cdots n}$.
Summing over all possible $i$ leads to the piece
\bea
{\cal P}_9=-{\tau\over4}\,\Big(\prod_{j=2}^{n-1}\,k_j\cdot X^{(0)}_j\Big)\,\Big[\Big((k_a-k_b)\cdot\partial_{k_1}+(k_b-k_a)\cdot\partial_{k_n}\Big)\,{\rm A}(2\shuffle\cdots\shuffle n-1)\Big]\,.~~\label{p9}
\eea
The operator $
(k_a-k_b)\cdot\partial_{k_1}+(k_b-k_a)\cdot\partial_{k_n}$
is introduced for two reasons. First, this operator is consistent with momentum conservation. Secondly, this operator annihilates terms from Feynman diagrams in
Figure.\ref{eli}  which also contribute to ${\rm A}(2\shuffle\cdots\shuffle n-1)$, leaves only desired contributions correspond to the configuration in Figure.\ref{abYm}.

\begin{figure}
  \centering
  \includegraphics[width=8cm]{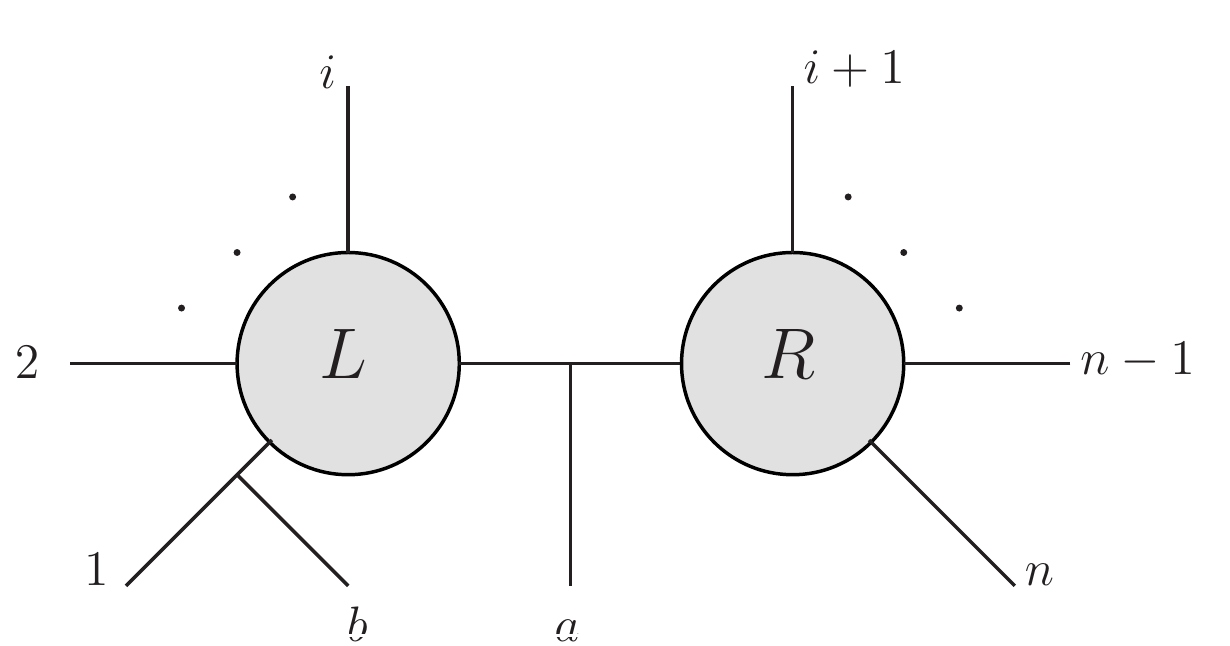}
   \includegraphics[width=8cm]{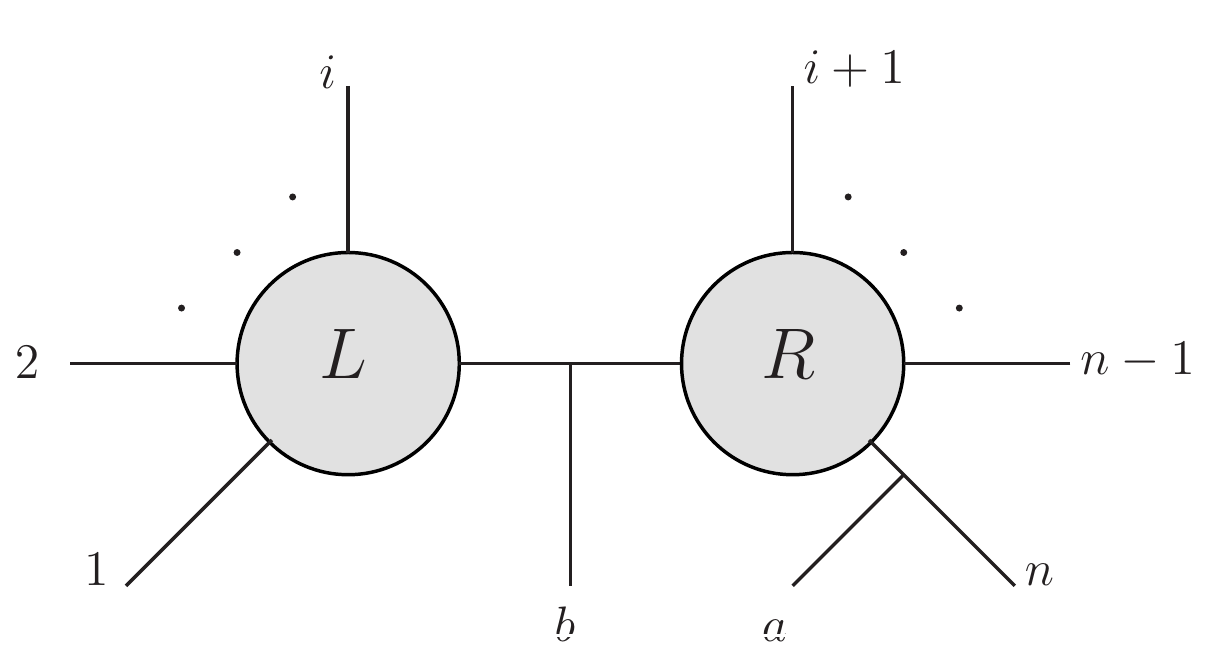}  \\
  \caption{Fifth type of Feynman diagrams.}\label{abIIm}
\end{figure}

Finally, we should consider diagrams provided in Figure.\ref{abIIm}. Let us focus on the first configuration in Figure.\ref{abIIm}.
At the r.h.s of \eref{expan-for-ds}, the corresponding terms are
\bea
(k_b\cdot k_1)\,(k_a\cdot K_{b1\cdots i})\,\Big(\prod_{j=2}^{n-1}\,k_j\cdot X_j\Big)\,{\rm A}(1,b,2\shuffle\cdots\shuffle i,a,i+1\shuffle\cdots\shuffle n-1)\,.
\eea
When $k_a\to\tau k_a,\,k_b\to\tau k_b$, leading contributions are at the $\tau^1$ order and are given by
\bea
& &-\tau\,{k_a\cdot K_{1\cdots i}\over 2s_{1\cdots i}^2}\,\Big(\prod_{j=2}^{n-1}\,k_j\cdot X^{(0)}_j\Big)\,{\cal N}_L\,{\cal N}_R\nn
&=&{\tau\over4}\,\Big(k_a\cdot\partial_{k_1}\,{1\over s_{1\cdots i}}\Big)\,\Big(\prod_{j=2}^{n-1}\,k_j\cdot X^{(0)}_j\Big)\,{\cal N}_L\,{\cal N}_R\,.
\eea
Summing over all possible $i$, we get
\bea
{\cal P}_{10;1}={\tau\over4}\,\Big(\prod_{j=2}^{n-1}\,k_j\cdot X^{(0)}_j\Big)\,\Big(k_a\cdot\partial_{k_1}\,{\rm A}(2\shuffle\cdots\shuffle n-1)\Big)\,.~~\label{p101}
\eea
Applying the same manipulation to the second configuration in Figure.\ref{abIIm} gives
\bea
{\cal P}_{10;2}={\tau\over4}\,\Big(\prod_{j=2}^{n-1}\,k_j\cdot X^{(0)}_j\Big)\,\Big(k_b\cdot\partial_{k_n}\,{\rm A}(2\shuffle\cdots\shuffle n-1)\Big)\,.~~\label{p102}
\eea
Again, the combination ${\cal P}_{10;1}+{\cal P}_{10;2}$ is consistent with momentum conservation while individual pieces are not.

Now we are ready to figure out the sub-leading double soft operator. To realize this, we need to regroup the sub-leading contributions at the $\tau^1$ order as
\bea
{\cal A}^{(1)}_{\rm N}(1,\cdots,n,a,b)=S^{(1)}_{\rm N}(a,b)\,{\cal A}_{\rm N}(1,\cdots,n)\,,
\eea
where $S^{(1)}_{\rm N}(a,b)$ is an operator at the $\tau^1$ order.
We first add ${\cal P}_{2;1}$, ${\cal P}_{5;1}$,
${\cal P}_{6;1}$, ${\cal P}_{2;2}$, ${\cal P}_{5;2}$,
${\cal P}_{6;2}$ in \eref{p21}, \eref{p51}, \eref{p61}, \eref{p22}, \eref{p52} and \eref{p62} together to get
\bea
{\cal R}_1=-\tau\,(k_a\cdot k_b)\,\Big({k_a\cdot k_1\over 4\big((k_a+k_b)\cdot k_1\big)^2}+{k_b\cdot k_n\over 4\big((k_a+k_b)\cdot k_n\big)^2}\Big)\,{\cal A}_{\rm N}(1,\cdots,n)\,.~~\label{R1}
\eea
Then, by combining ${\cal P}_{3;1}$, ${\cal P}_{7;1}$, ${\cal P}_{3;2}$, ${\cal P}_{7;2}$, ${\cal P}_8$ in \eref{p31}, \eref{p71},
\eref{p32}, \eref{p72} and \eref{p8} together, we obtain
\bea
{\cal R}_{21}&=&\tau\,\Big[\,{\cal J}
\Big(\prod_{i=2}^{n-1}\,k_i\cdot X^{(0)}_i\Big)\Big]\,{\rm A}(2\shuffle\cdots\shuffle n-1)\,,~~\label{R21}
\eea
where
\bea
{\cal J}&=&{(k_a\cdot k_1)\,(k_b\cdot\partial_{k_1})-(k_b\cdot k_1)\,(k_a\cdot\partial_{k_1})\over 4(k_a+k_b)\cdot k_1}
+{(k_b\cdot k_n)\,(k_a\cdot\partial_{k_n})-(k_a\cdot k_n)\,(k_b\cdot\partial_{k_n})\over 4(k_a+k_b)\cdot k_n}\nn
&=&{k_a\cdot J_1\cdot k_b\over 4(k_a+k_b)\cdot k_1}+{k_b\cdot J_n\cdot k_a\over 4(k_a+k_b)\cdot k_n}\,.
\eea
In the second line at the r.h.s, $J^{\mu\nu}_a$ is the angular momentum operator of the scalar particle $a$ which is defined by
\bea
J_a^{\mu\nu}\equiv k_a^\mu{\partial\over\partial k_{a,\nu}}-k_a^\nu{\partial\over\partial k_{a,\mu}}\,.
\eea
Putting pieces ${\cal P}_{1;1}$, ${\cal P}_{4;1}$, ${\cal P}_{10;1}$,
${\cal P}_{1;2}$, ${\cal P}_{4;2}$, ${\cal P}_{10;2}$, ${\cal P}_9$ in \eref{p11}, \eref{p41}, \eref{p101}, \eref{p12}, \eref{p42}, \eref{p102},
\eref{p9} together, we find
\bea
{\cal R}_{22}=\tau\,\Big(\prod_{i=2}^{n-1}\,k_i\cdot X^{(0)}_i\Big)\,\Big({\cal J}\,{\rm A}(2\shuffle\cdots\shuffle n-1)\Big)\,.~~\label{R22}
\eea
One can combine ${\cal R}_{21}$ and ${\cal R}_{22}$ to get
\bea
{\cal R}_2={\cal R}_{21}+{\cal R}_{22}=\tau\,{\cal J}\,{\cal A}_{\rm N}(1,\cdots,n)\,.
\eea
Consequently, we can express ${\cal A}^{(1)}_{\rm N}(1,\cdots,n,a,b)$ as
\bea
{\cal A}^{(1)}_{\rm N}(1,\cdots,n,a,b)={\cal R}_1+{\cal R}_2=S^{(1)}_{\rm N}(a,b)\,{\cal A}^{(1)}_{\rm N}(1,\cdots,n)\,,
\eea
where the soft operator $S^{(1)}_{\rm N}(a,b)$ is given as
\bea
S^{(1)}_{\rm N}(a,b)&=&-\tau\,\Big[(k_a\cdot k_b)\,\Big({k_a\cdot k_1\over 4\big((k_a+k_b)\cdot k_1\big)^2}+{k_b\cdot k_n\over 4\big((k_a+k_b)\cdot k_n\big)^2}\Big)\nn
& &+{k_b\cdot J_1\cdot k_a\over 4(k_a+k_b)\cdot k_1}+{k_a\cdot J_n\cdot k_b\over 4(k_a+k_b)\cdot k_n}\Big]\,,
\eea
which is again the same as the result found in \cite{Cachazo:2015ksa,Du:2015esa}.

\section{Summery}
\label{sec-conclu}

In this note, we determined tree NLSM amplitudes by assuming the universality of single soft behavior. The assumption of double copy structure was also used, which tells us coefficients ${\cal C}(\sigma'_{n-2},k_i)$ in \eref{exp-N-KK} are independent of the ordering $\sigma_{n-2}$. We first observed the Adler's zero for $4$-point NLSM amplitudes, by considering kinematics. Then, we figured out the expanded formula of general tree NLSM amplitudes, which manifests the permutation invariance among external legs, by using the universality of the Adler's zero. Notice again the Adler's zero was not assumed at the beginning, it serves as a deduced property. We also re-derived double soft factors for tree NLSM amplitudes at leading and sub-leading orders, via the resulting expanded formula. The obtained double soft factors are coincide with those in literatures.

Our soft bootstrap method, based on universality of soft behaviors and double copy structure, are proven to be useful for constructing tree amplitudes of Yang-Mills-scalar, Yang-Mills, Einstein-Yang-Mills, gravity \cite{Zhou:2022orv}, as well as non-linear sigma model. It is interesting to apply this method to a wider range of theories. Another potential future direction is to generalize this method from the tree level
to the loop level.

\section*{Acknowledgments}

It is our great pleasure to thank Prof.Yijian Du for useful discussions and comments.



\begin{thebibliography}{}
\bibliographystyle{JHEP}

\bibitem{Cachazo:2014fwa}
F.~Cachazo and A.~Strominger,
``Evidence for a New Soft Graviton Theorem,''
[arXiv:1404.4091 [hep-th]].

\bibitem{Schwab:2014xua}
B.~U.~W.~Schwab and A.~Volovich,
``Subleading Soft Theorem in Arbitrary Dimensions from Scattering Equations,''
Phys. Rev. Lett. \textbf{113}, no.10, 101601 (2014)
doi:10.1103/PhysRevLett.113.101601
[arXiv:1404.7749 [hep-th]].

\bibitem{Afkhami-Jeddi:2014fia}
N.~Afkhami-Jeddi,
``Soft Graviton Theorem in Arbitrary Dimensions,''
[arXiv:1405.3533 [hep-th]].


\bibitem{Cheung:2014dqa}
C.~Cheung, K.~Kampf, J.~Novotny and J.~Trnka,
``Effective Field Theories from Soft Limits of Scattering Amplitudes,''
Phys. Rev. Lett. \textbf{114}, no.22, 221602 (2015)
doi:10.1103/PhysRevLett.114.221602
[arXiv:1412.4095 [hep-th]].

\bibitem{Luo:2015tat}
H.~Luo and C.~Wen,
``Recursion relations from soft theorems,''
JHEP \textbf{03}, 088 (2016)
doi:10.1007/JHEP03(2016)088
[arXiv:1512.06801 [hep-th]].

\bibitem{Cheung:2018oki}
C.~Cheung, K.~Kampf, J.~Novotny, C.~H.~Shen, J.~Trnka and C.~Wen,
Phys. Rev. Lett. \textbf{120}, no.26, 261602 (2018)
doi:10.1103/PhysRevLett.120.261602
[arXiv:1801.01496 [hep-th]].

\bibitem{Elvang:2018dco}
H.~Elvang, M.~Hadjiantonis, C.~R.~T.~Jones and S.~Paranjape,
``Soft Bootstrap and Supersymmetry,''
JHEP \textbf{01}, 195 (2019)
doi:10.1007/JHEP01(2019)195
[arXiv:1806.06079 [hep-th]].

\bibitem{Cachazo:2016njl}
F.~Cachazo, P.~Cha and S.~Mizera,
``Extensions of Theories from Soft Limits,''
JHEP \textbf{06}, 170 (2016)
doi:10.1007/JHEP06(2016)170
[arXiv:1604.03893 [hep-th]].


\bibitem{Rodina:2018pcb}
L.~Rodina,
``Scattering Amplitudes from Soft Theorems and Infrared Behavior,''
Phys. Rev. Lett. \textbf{122}, no.7, 071601 (2019)
doi:10.1103/PhysRevLett.122.071601
[arXiv:1807.09738 [hep-th]].

\bibitem{Boucher-Veronneau:2011rwd}
C.~Boucher-Veronneau and A.~J.~Larkoski,
``Constructing Amplitudes from Their Soft Limits,''
JHEP \textbf{09}, 130 (2011)
doi:10.1007/JHEP09(2011)130
[arXiv:1108.5385 [hep-th]].

\bibitem{Nguyen:2009jk}
D.~Nguyen, M.~Spradlin, A.~Volovich and C.~Wen,
``The Tree Formula for MHV Graviton Amplitudes,''
JHEP \textbf{07}, 045 (2010)
doi:10.1007/JHEP07(2010)045
[arXiv:0907.2276 [hep-th]].


\bibitem{Low}
  F.~E.~Low,
  ``Bremsstrahlung of very low-energy quanta in elementary particle collisions,''
  Phys.\ Rev.\  {\bf 110}, 974 (1958).

\bibitem{Weinberg}
 S.~Weinberg,
  ``Infrared photons and gravitons,''
  Phys.\ Rev.\  {\bf 140}, B516 (1965).


\bibitem{Casali:2014xpa}
E.~Casali,
``Soft sub-leading divergences in Yang-Mills amplitudes,''
JHEP \textbf{08}, 077 (2014)
doi:10.1007/JHEP08(2014)077
[arXiv:1404.5551 [hep-th]].

\bibitem{Britto:2004ap}
R.~Britto, F.~Cachazo and B.~Feng,
``New recursion relations for tree amplitudes of gluons,''
Nucl. Phys. B \textbf{715}, 499-522 (2005)
doi:10.1016/j.nuclphysb.2005.02.030
[arXiv:hep-th/0412308 [hep-th]].

\bibitem{Britto:2005fq}
R.~Britto, F.~Cachazo, B.~Feng and E.~Witten,
``Direct proof of tree-level recursion relation in Yang-Mills theory,''
Phys. Rev. Lett. \textbf{94}, 181602 (2005)
doi:10.1103/PhysRevLett.94.181602
[arXiv:hep-th/0501052 [hep-th]].


\bibitem{Cachazo:2013gna}
F.~Cachazo, S.~He, and E.~Y. Yuan,
``{Scattering Equations and Kawai-Lewellen-Tye Orthogonality},''
 { Phys. Rev.} {\bf
  D90} (2014) no. 6, 065001,
{ arXiv:1306.6575 [hep-th]}.

\bibitem{Cachazo:2013hca}
F.~Cachazo, S.~He, and E.~Y. Yuan,
``{Scattering of Massless Particles in Arbitrary Dimensions},''
  { Phys. Rev.
  Lett.} {\bf 113} (2014) no. 17, 171601,
 arXiv:1307.2199 [hep-th].

\bibitem{Cachazo:2013iea}
F.~Cachazo, S.~He, and E.~Y. Yuan,
``{Scattering of Massless Particles: Scalars, Gluons and Gravitons},''
  { JHEP} {\bf 1407} (2014)
   033,
 arXiv:1309.0885 [hep-th].

\bibitem{Cachazo:2014nsa}
  F.~Cachazo, S.~He and E.~Y.~Yuan,
  ``Einstein-Yang-Mills Scattering Amplitudes From Scattering Equations,''
  JHEP {\bf 1501}, 121 (2015)
  [arXiv:1409.8256 [hep-th]].


\bibitem{Cachazo:2014xea}
  F.~Cachazo, S.~He and E.~Y.~Yuan,
  ``Scattering Equations and Matrices: From Einstein To Yang-Mills, DBI and NLSM,''
  JHEP {\bf 1507}, 149 (2015)
  [arXiv:1412.3479 [hep-th]].


\bibitem{Zhou:2022orv}
K.~Zhou,
``Tree level amplitudes from soft theorems,''
JHEP \textbf{03}, 021 (2023)
doi:10.1007/JHEP03(2023)021
[arXiv:2212.12892 [hep-th]].


\bibitem{Kawai:1985xq}
  H.~Kawai, D.~C.~Lewellen and S.~H.~Tye,
  ``A Relation Between Tree Amplitudes of Closed and Open Strings,''
  Nucl.\ Phys.\ B {\bf 269}, 1 (1986).

\bibitem{Bern:2008qj}
  Z.~Bern, J.~J.~M.~Carrasco and H.~Johansson,
  ``New Relations for Gauge-Theory Amplitudes,''
  Phys.\ Rev.\ D {\bf 78}, 085011 (2008)
  [arXiv:0805.3993 [hep-ph]].

\bibitem{Chiodaroli:2014xia}
  M.~Chiodaroli, M.~Gnaydin, H.~Johansson and R.~Roiban,
  ``Scattering amplitudes in $ \mathcal{N}=2 $ Maxwell-Einstein and Yang-Mills/Einstein supergravity,''
  JHEP {\bf 1501}, 081 (2015)
  doi:10.1007/JHEP01(2015)081
  [arXiv:1408.0764 [hep-th]].

\bibitem{Johansson:2015oia}
  H.~Johansson and A.~Ochirov,
  ``Color-Kinematics Duality for QCD Amplitudes,''
  JHEP {\bf 1601}, 170 (2016)
  doi:10.1007/JHEP01(2016)170
  [arXiv:1507.00332 [hep-ph]].

\bibitem{Johansson:2019dnu}
  H.~Johansson and A.~Ochirov,
  ``Double copy for massive quantum particles with spin,''
  JHEP {\bf 1909}, 040 (2019)
  doi:10.1007/JHEP09(2019)040
  [arXiv:1906.12292 [hep-th]].


\bibitem{Adler}
S.~L.~Adler,
''Consistency conditions on the strong interactions implied by a partially conserved axial vector
current,''
Phys.\ Rev. 137 (1965) B1022–B1033


\bibitem{Cachazo:2015ksa}
F.~Cachazo, S.~He and E.~Y.~Yuan,
Phys. Rev. D \textbf{92}, no.6, 065030 (2015)
doi:10.1103/PhysRevD.92.065030
[arXiv:1503.04816 [hep-th]].

\bibitem{Du:2015esa}
Y.~J.~Du and H.~Luo,
JHEP \textbf{08}, 058 (2015)
doi:10.1007/JHEP08(2015)058
[arXiv:1505.04411 [hep-th]].


\bibitem{Kleiss:1988ne}
  R.~Kleiss and H.~Kuijf,
  ``MULTI - GLUON CROSS-SECTIONS AND FIVE JET PRODUCTION AT HADRON COLLIDERS,''
  Nucl.\ Phys.\  B {\bf 312}, 616 (1989).


\end{thebibliography}
\end{document}